# Metastable defects decrease the fill factor of solar cells

Thomas Paul Weiss[1], Omar Ramírez[1], Stefan Paetel[2], Wolfram Witte[2], Jiro Nishinaga[3], Thomas Feurer[4], Susanne Siebentritt[1]

[1] University of Luxembourg, Department of Physics and Materials Science, 41, rue du Brill, L-4422 Belvaux, Luxembourg

[2] Zentrum für Sonnenenergie- und Wasserstoff-Forschung Baden-Württemberg (ZSW), Meitnerstraße 1, 70563 Stuttgart, Germany

[3] Research Institute for Energy Conservation, National Institute of Advanced Industrial Science and Technology (AIST), 1-1-1 Umezono, Tsukuba, Japan

[4] Laboratory for Thin Films and Photovoltaics, Empa - Swiss Federal Laboratories for Materials Science and Technology, Überlandstraße 129, 8600 Dübendorf, Switzerland

**Abstract**

**$Cu(In,Ga)Se_2$ based solar cells exceed power conversion efficiencies of 23 %. Yet, the fill factor of these solar cells, with best values around 80 %, is relatively low (Si reaches 84.9%) mostly due to diode factors greater than one. Recently, we proposed metastable defects, a general feature of the $Cu(In,Ga)Se_2$ alloy, to be the origin of the increased diode factor. We measure the diode factor of the bare absorber layers by excitation-dependent photoluminescence. For high quality and thus high luminescent polycrystalline absorbers, we evaluate the diode factor excitation dependence over four orders of magnitude. Using simulations and the model of metastable defects, we can well describe the experimental findings on n- and p-type epitaxial films as well as the polycrystalline absorbers, providing additional evidence for this model. We find that the diode factors measured optically by photoluminescence impose a lower limit for the diode factor measured electrically on a finished solar cell. Interestingly, the lowest diode factor (optical and electrical) and consequently highest fill factor of 81.0 % is obtained by Ag alloying, i.e. an $(Ag,Cu)(In,Ga)Se_2$ absorber. This finding hints to a pathway to increase fill factors and thus efficiencies for $Cu(In,Ga)Se_2$-based solar cells.**

# 1. Introduction

$Cu(In,Ga)Se_2$ (CIGS) is a material system used for high efficiency thin film photovoltaics. Record efficiencies in the laboratory reached already 23.35 % [1]. These solar cells are limited by non-radiative recombination channels [2], which was recently linked to grain boundary recombination resulting in the hypothesis that grain size needs to be increased for higher open-circuit voltages [3]. However, apart from voltage losses, state-of-the-art CIGS solar cells also suffer from relatively low fill factors (FF); the record CIGS solar cell has a FF around 80.4 % [1] compared to Si with a FF of 84.9 % [4, 5] at similar open-circuit voltages. These FF losses can be associated due to diode factors greater than one [6, 7]. In fact, already the diode factor of the bare absorber exceeds the theoretical value of one [7, 8] and thus raises the question on the origin of this loss mechanism.

We have recently shown that an increase in the net-doping, i.e. a downshift of the majority carrier Fermi level, upon illumination in low-injection conditions can cause the increase of the diode factor. The reason is a metastable change of the net acceptor density due to injection of free electrons [9], which is a universal feature of p-type CIGS semiconducting alloys [10]. The metastable increase of the net-acceptor density can be explained by the $V_{Se}$-$V_{Cu}$ defect complex [11], which well describes many experimental observations of metastable acceptor densities [7, 12, 13]. The defect complex involves large lattice relaxations, which can be expressed as the In-In atomic distance. For small and large In-In distances, the defect complex is in a donor and acceptor configuration, respectively. Due to the large-lattice relaxation, energy barriers are present for the transitions between donor and acceptor state, which is the reason of the metastability [14]. The exact nature of the metastable defect is not critical here. We use a model of a metastable defect that is in a donor state in equilibrium in a p-type material and changes to an acceptor state upon electron injection.

In reference [7] we used the model of such a metastable defect to explain the increased diode factor, which directly results in reduced fill factors. Due to the direct impact of the metastability on device performance it is thus of major importance to study the capabilities of the model to correctly describe experimental data. In particular, in reference [7] the simulated diode factor shows a clear dependence with respect to the excitation level within the model of metastable defects, i.e. it depends on the density of injected minority carriers (electrons for p-type CIGS). It is shown here that the diode factor measured experimentally shows exactly such excitation dependence.

The paper is structured as follows. First, the theoretical background of the metastable defect model is elaborated and tested on how the parameters of the metastable defect influence the excitation dependence of the diode factor. Next, epitaxial CIGS films are investigated, which are known to exist as n-type or p-type semiconducting layers [15, 16]. The results give additional support for the model of the $V_{Se}$-$V_{Cu}$ defect complex as the origin of the metastable increase of the net acceptor density. Finally, diode factors are measured optically on high efficiency CIGS from different laboratories. The weak excitation dependence of the diode factor is fitted with the model of metastable defects, which shows a good agreement. A comparison with a large set of solar cells from several different institutes show that the fill factor, and consequently the efficiency, is limited by the diode factor of state-of-the-art devices.

## 2. Theoretical background

As shown previously, the diode factor $A$ of the bare absorber can be measured and simulated using intensity-dependent photoluminescence (PL) [8] and expresses as

$$A = \frac{\partial \ln Y_{PL}}{\partial \ln G} \tag{1}$$

where $Y_{PL}$ is the integrated PL yield (measured or simulated, see section 3 for details) and $G$ the generation flux [7, 8, 17].

CIGS is a p-type absorber layer with dominating Shockley-Read-Hall (SRH) recombination, which is operated in low-injection conditions during our PL measurements. Therefore, a diode factor of 1 is expected, as only the electron Fermi level moves upon illumination [7, 18]. Recently, we have shown that a diode factor above 1 can be explained by an additional downshift of the hole quasi Fermi level upon illumination, which happens even in low-injection conditions [7]. We have shown that the downshift of the hole (majority) quasi Fermi level can be described by metastable defects involving large lattice relaxations [7]. In CIGS, the $V_{Cu}$-$V_{Se}$ divacancy complex is such a defect [11], which has a donor and an acceptor state, whose occupancy in the acceptor state $f_A$ depends on the rate constants of the defect transitions and is described by (2) [13].

$$f_A = \frac{\tau_{\mathrm{HE}}^{-1} + \tau_{EC}^{-1}}{\tau_{HE}^{-1} + \tau_{EC}^{-1} + \tau_{EE}^{-1} + \tau_{HC}^{-1}} \tag{2}$$

In (2), the $\tau_{ij}^{-1}$ describe the rate constants for the most probable transitions [11] for the conversion between the donor and the acceptor state, with i = E or H abbreviating electron or hole, and j = E or C abbreviating emission or capture. Each transition process with the rate constant $\tau_{ij}^{-1}$ involves two charge carriers. Expressions for $\tau_{ij}^{-1}$ and their transition processes are summarized in Table 1. The occupancy in the donor state is given by $1 - f_A$. We should note that we use the $V_{Se}$-$V_{Cu}$ divacancy as the model for the metastable defect. However, any metastable defect has a forward and a backward reaction governed by rate constants and activation energies. Thus, the model is not specific to the $V_{Se}$-$V_{Cu}$ divacancy.

Table 1 – Summary for the dominating processes governing the transition of the metastable defect with large lattice relaxations. Here, the expressions are explicitly written for the $V_{Se}$-$V_{Cu}$ divacancy defect in CIGS.

| Rate constant | Process name | Process description | Expression | Transition process |
|---|---|---|---|---|
| $\tau_{EC}^{-1}$ | 'Electron capture' | Electron capture + hole emission | $\tau_{EC}^{-1} = nN_V\, exp\left(-\frac{\Delta E_{EC}}{k_B T}\right)$ | Donor → Acceptor |
| $\tau_{EE}^{-1}$ | 'Electron emission' | Electron emission + hole capture | $\tau_{EE}^{-1} = pN_C\, exp\left(-\frac{\Delta E_{EE}}{k_B T}\right)$ | Acceptor → Donor |
| $\tau_{HC}^{-1}$ | 'Hole capture' | Double hole capture | $\tau_{HC}^{-1} = p^2\, exp\left(-\frac{\Delta E_{HC}}{k_B T}\right)$ | Acceptor → Donor |
| $\tau_{HE}^{-1}$ | 'Hole emission' | Double hole emission | $\tau_{HE}^{-1} = N_V^2\, exp\left(-\frac{\Delta E_{HE}}{k_B T}\right)$ | Donor → Acceptor |

Using the expressions for the transition rates $\tau_{ij}^{-1}$, $f_A$ in eqn. (2) expresses as

$$f_A = \frac{N_V^2\, exp\left(-\frac{\Delta E_{HE}}{k_B T}\right) + nN_V\, exp\left(-\frac{\Delta E_{EC}}{k_B T}\right)}{N_V^2\, exp\left(-\frac{\Delta E_{HE}}{k_B T}\right) + nN_V\, exp\left(-\frac{\Delta E_{EC}}{k_B T}\right) + pN_C\, exp\left(-\frac{\Delta E_{EE}}{k_B T}\right) + p^2\, exp\left(-\frac{\Delta E_{HC}}{k_B T}\right)} \quad (3)$$

From (3) it becomes obvious that the occupancy depends on the minority carrier density and therefore on the excitation conditions, i.e. the generation flux $G$. In low-injection conditions and with reasonable good transport properties (flat quasi Fermi levels), the minority carrier density $n$ is given by

$$n \approx G\tau_n/d \quad (4)$$

where $\tau_n$ denotes the minority carrier lifetime and $d$ the absorber thickness. The hole density in (3) is given in the low injection case by $p \approx N_A + f_A N_{MS} - (1 - f_A)N_{MS}$, which is the fixed (non-metastable) acceptor density $N_A$ plus the metastable defects in the acceptor configuration $f_A N_{MS}$ minus the metastable defects in the donor configuration $(1 - f_A)N_{MS}$. Finally, the diode factor can be calculated as follows (see ref. [7] for details)

1. Determine the quasi Fermi levels $E_{fn}$ and $E_{fp}$ for electrons and holes respectively using the charge neutrality condition and the condition that generation equals recombination in steady state conditions. Eqn. (3) is taken into account and solved self consistently, i.e. the occupancy of the metastable defect is respected.
2. Calculate the quasi Fermi level splitting $\Delta\mu = E_{fn} - E_{fp}$ from which the relative PL yield is obtained according to Planck's generalized law [19]. It is noted that only the relative PL is calculated according to $Y_{PL} \propto \exp(\Delta\mu/k_B T)$. This expression is sufficient here as we only compare the resulting diode factors given by (1). In (1), any proportionality factor drops out due to the derivative of the logarithm.

3. Calculate the diode factor using eqn. (1).

The simulations in Figure 1 exemplify this process for the curve labeled $A = 1.40$ focusing on low-injection conditions. Details of Figure 1 are discussed below. With increasing generation flux, $E_{fn}$ increases as expected from eqn. (4). In the case of SRH recombination without metastable defects, $E_{fp}$ is constant (dash dotted line) and the shift of $E_{fn}$ results in $A = 1$. However, with the involvement of metastable defects, $E_{fp}$ shifts towards the valence band due to their conversion from donor to acceptor state. This additional shift results in $A > 1$. The calculated relative PL yields are shown in Figure 1b. The derivative in a double-logarithmic plot (eqn. (1)) gives the diode factor shown in Figure 1c.

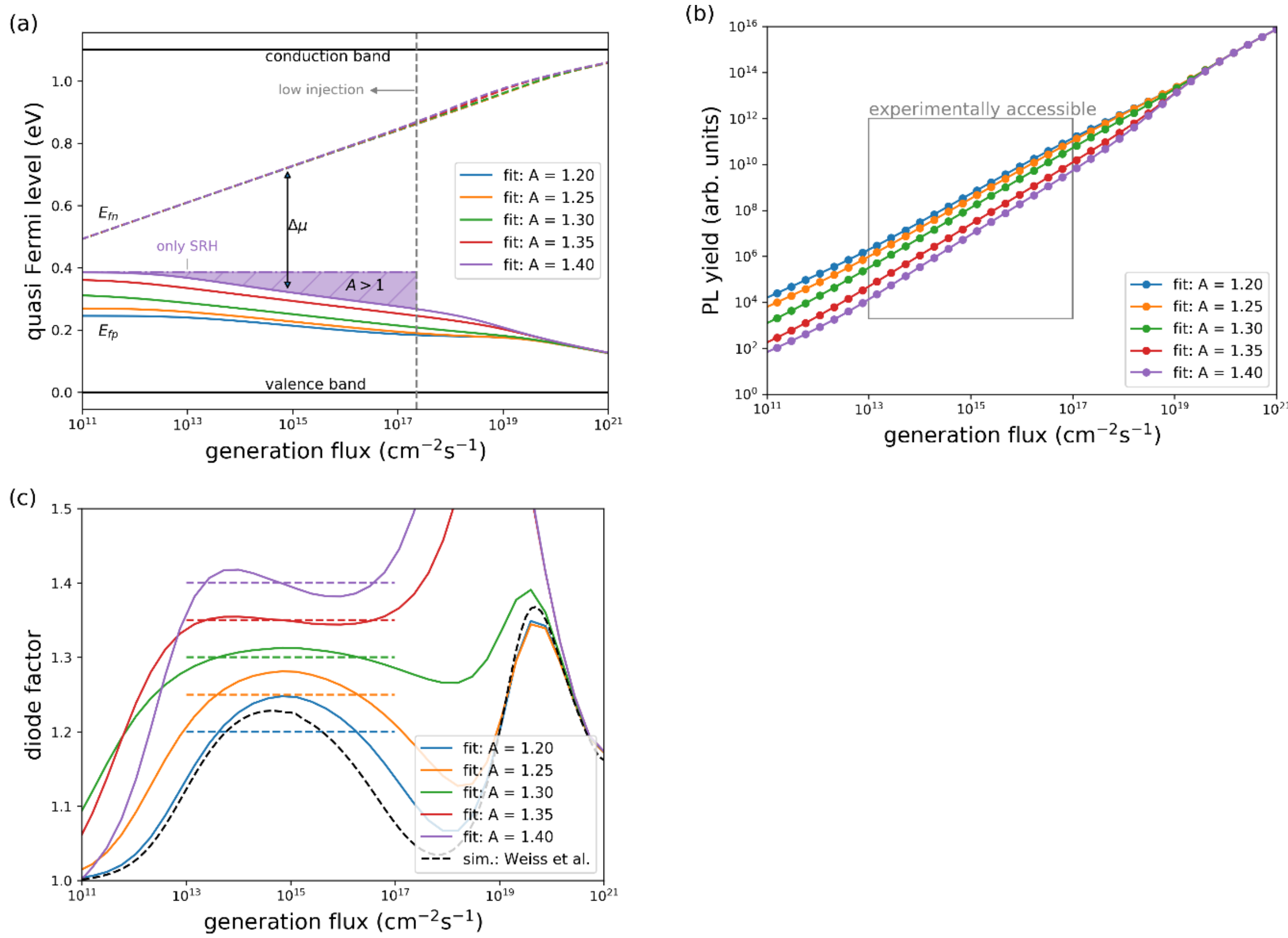

Figure 1 – a) Calculated electron and hole quasi Fermi levels for a semiconducting thin film including metastable defects. Due to the conversion of metastable defects from donor to acceptor configuration the hole quasi Fermi level shifts towards the valence band even in low-injection conditions. b) Calculated relative PL yield based on the quasi Fermi level splitting $\Delta\mu$ from (a). c) Calculated diode factor as a function of the generation flux using eqn. (1). Parameters for the calculations are determined by fitting to minimize the residuals to a constant diode factor within the range of generation fluxes of $10^{13}$ and $10^{17}$ $cm^{-2}s^{-1}$ (gray dashed lines). Black dashed line in (c) shows the simulation reported previously [7].

Experimentally, the diode factor is often not determined differentially according to eqn. (1), i.e. as a function of the generation flux, but from a straight line fit in a double logarithmic plot, see e.g. [7, 8, 18, 20]. Using such a linear fit instead of eqn. (1) assumes directly a generation independent diode factor and thus seems to be incompatible with the model of metastable defects. Indeed, upon closer inspection, the PL yield has generally a negative curvature with respect to the generation flux [7]. For the CIGS samples studied here, this is evidenced by the U-shaped residuals of the experimental PL yield and a straight line fit (Supplementary Figure 7).

As will be shown in the experimental section, high quality CIGS absorbers exhibit a diode factor greater than one with a small or even no generation flux dependence. Here, we demonstrate the capabilities of the metastable defect model and under which circumstances an almost generation flux independent diode factor greater than one can be obtained. We utilize a fitting routine, to determine the parameters of the metastable defects in order to obtain an approximately constant diode factor for generation fluxes between $10^{13}$ and $10^{17}$ $cm^{-2}s^{-1}$. This range is the experimentally accessible range to measure the PL yield (and thus the diode factor).

Prior to simply applying eqn. (3), which is needed for the calculation of the quasi Fermi-level splitting and thus for the fitting of the diode factor, the following considerations and parameter choices are made, which facilitate the fitting routine.

- The bandgap and the Fermi level dictate the equilibrium densities $n_0$ and $p_0$ of free carriers in the conduction and valence band, respectively, and in turn define the equilibrium occupancy $f_A$ via eqn. (3). The position of the electron Fermi level, where $f_A = 0.5$ is defined as the charge transition level $E_{TR}$. From detailed balance considerations, $E_g$, $E_{TR}$, and the four $\Delta E_{ij}$ are not independent from each other but obey the following two relations [13]

$$E_{TR} = \frac{1}{2}\left(E_g + \Delta E_{EC} - \Delta E_{EE}\right) \tag{5}$$

$$E_{TR} = \frac{1}{2}(\Delta E_{HE} - \Delta E_{HC}) \tag{6}$$

  Here, zero energy was set to $E_V = 0$. In the following discussion and simulations, $E_g$ is fixed to 1.1 eV. Then, $E_{TR}$ and $\Delta E_{HE}$ are calculated via eqn. (5) and (6), respectively, by using the remaining three free parameters $\Delta E_{EC}$, $\Delta E_{EE}$, and $\Delta E_{HC}$.

- From eqn. (3) it is obvious that the occupancy $f_A$ depends not on the absolute values of $\Delta E_{ij}$, i.e. $f_A$ is invariant upon shifting all $\Delta E_{ij}$ by the same constant $\Delta E$. This additional degree of freedom allows to fix one of the remaining energy barriers ($\Delta E_{EC}$, $\Delta E_{EE}$, and $\Delta E_{HC}$). While this procedure does not change $f_A$, it implies different dynamics of the metastable defect. However, the dynamics are not part of the present paper. Consequently, if not mentioned otherwise, we fix $\Delta E_{EC} = 0.35$ eV without loss of generality.

- The non-metastable doping density $N_A$ is fixed to $10^{16}$ cm$^{-3}$, while the density of metastable defects $N_{MS}$ is allowed to vary during the fitting of the diode factor. Similar simulation results are obtained for fixed doping densities for instance of $10^{17}$ cm$^{-3}$ (Supplementary Figure 2), as mainly the difference between the doping density and metastable defect density is of importance [7].

To check the validity of the simulations, the transition rates from acceptor to donor $U_{A\to D}$ and from donor to acceptor $U_{D\to A}$ are calculated according to [13]

$$U_{A\to D} = f_A(\tau_{EE}^{-1} + \tau_{HC}^{-1})N_t$$

$$U_{D\to A} = (1 - f_A)(\tau_{HE}^{-1} + \tau_{EC}^{-1})N_t$$

For the solution at equilibrium, these rates have to be equal (see Supplementary Figure 9).

Figure 1c shows the simulations (solid lines), where the diode factor is fitted to various generation flux independent diode factors between 1.2 and 1.4. Parameters for the metastable defect are listed in Table 2. As described above, the reason for diode factors above one is the down-shift of the hole quasi Fermi level as shown in Figure 1a as a consequence of the conversion of metastable defects from donor to acceptor state. It is noted that low-injection conditions prevail in these simulations (Supplementary Figure 8) for generation fluxes below $\approx 3 \cdot 10^{17}$ cm$^{-2}$s$^{-1}$.

It is interesting to note that the best fitting results are obtained for diode factors between 1.30 and 1.35. For smaller diode factors, the generation flux dependence increases within the model of one metastable defect. Supplementary section A gives further details concerning the dependence of the rate constants $\tau_{ij}^{-1}$ as a function of the free carrier densities, which change as a function of the generation flux.

Table 2 – Fitted parameters for the metastable defects to describe the generation flux dependence of the diode factor shown in Figure 1c and Figure 3b. The barrier for the electron capture was set constant to $\Delta E_{EC} = 0.35\ eV$ and the fixed doping density to $N_A = 10^{16}\ cm^{-3}$. The experimental data shown in Figure 3b are fitted with the boundary condition of a free hole density of $p_{0,bound} = 10^{14} cm^{-3}$. Only for the Empa sample a significant better fit is obtained with the relaxed boundary condition of $p_{0,bound} = 10^{13} cm^{-3}$ and thus these values are given here as well.

| **Sample** | $\Delta E_{HC}$ **/ eV** | $\Delta E_{EE}$ **/ eV** | $N_{MS}$ **/ cm$^{-3}$** |
|---|---|---|---|
| | | | |
| **Figure 1a (fit to constant diode factor)** | | | |
| **A = 1.20** | 0.319 | 0.696 | $8.610 \cdot 10^{15}$ |
| **A = 1.25** | 0.278 | 0.654 | $9.462 \cdot 10^{15}$ |
| **A = 1.30** | 0.205 | 0.581 | $9.908 \cdot 10^{15}$ |
| **A = 1.35** | 0.082 | 0.354 | $9.977 \cdot 10^{15}$ |
| **A = 1.40** | 0.013 | 0.534 | $1.007 \cdot 10^{16}$ |
| **Figure 3b (fit to experimental data with $p_{0,bound} = 10^{14}cm\text{^} - 3$ )** | | | |
| **ZSW** | $0.22 \pm 0.02$ | $0.73 \pm 0.19$ | $10^{16.014 \pm 0.016}$ |
| **ZSW-Ag** | $0.34 \pm 0.01$ | $> 0.75$ | $10^{15.6 \pm 1.5}$ |
| **AIST** | $0.20 \pm 0.03$ | $0.76 \pm 0.06$ | $10^{16.06 \pm 0.04}$ |
| **Empa** | $0.17 \pm 0.03$ | $0.66 \pm 0.03$ | $10^{16.05 \pm 0.03}$ |
| **Empa ($p_{0,bound} = 10^{13}cm^{-3}$)** | $0.09 \pm 0.02$ | $0.56 \pm 0.03$ | $10^{16.013 \pm 0.005}$ |

# 3. Experimental

To support the theoretical model of metastable defects introduced in [7] and elaborated above, epitaxial and polycrystalline CIGS films are investigated. The epitaxial films under investigation here have neither grain boundaries nor extrinsically added doping, such as alkalis. Thus, the CIGS bulk properties are explored. High-quality polycrystalline CIGS films are used to support the metastable defect model by measurements of the diode factor on a sufficient wide range of generation fluxes. In addition, the relevance of the diode factor on the fill factor and the efficiency of finished solar cells is demonstrated.

Epitaxial Cu(In,Ga)$Se_2$ absorber were grown by metalorganic vapor phase epitaxy on 100-oriented undoped GaAs substrates at 520°C. In order to tune the Ga content, a 2-step growth process was implemented. First, a Cu(In,Ga)$Se_2$ layer with [Ga]/([Ga]+[In]) (GGI) of 0.4 is grown, followed by a layer of pure $CuInSe_2$. By adjusting the thickness of the first and second layer, a precise control of the integral Ga content of the final CIGS layer can be achieved. All samples have a [Cu]/([Ga]+[In]) ratio between 0.83 and 0.89 (Cu-poor), which is a necessary condition to obtain n-type conductivity [21]. Details of the sample's growth and the dependency of the conduction type (n-type or p-type) on the gallium content can be found in reference [16]. In particular, for compositional ratios of GGI $\geq$ 0.19, p-type conduction is observed, while for GGI $\leq$ 0.15, n-type conduction is obtained [16].

Polycrystalline CIGS absorber layers and their respective solar cell devices are grown by ZSW, AIST, and Empa. The absorber layers are grown following a three-stage or modified three-stage co-evaporation process [22, 23] from elemental evaporation sources. The deposition process results in a double graded compositional profile such that the Ga concentration has a minimum within the bulk of the absorber layer. As a result, the bandgap is graded with a minimum corresponding to the minimum in Ga concentration. Additional details concerning the growth process are given below.

- ZSW: A Mo-coated soda-lime glass served as substrate providing Na and partially K for diffusion into CIGS layer at elevated temperatures. Absorber layers are grown in a 30 x 30 $cm^2$ in-line deposition machine with a multi-stage co-evaporation process. For some absorber layers, also Ag is co-evaporated. At the end of the CIGS process, an *in-situ* RbF post-deposition treatment (PDT) is carried out under Se atmosphere [24, 25]. Solar cells are finished using a solution-grown

CdS buffer layer, a sputtered ZnO/Al:ZnO double window layer and a Ni/Al contact grid. Individual solar cells are scribed mechanically with an area of 0.5 $cm^2$. Solar cells yield efficiencies around 18.5 % (total area; without anti-reflective coating). The PL peak position, representative for the bandgap minimum [26], is around 1.10 – 1.15 eV. In total, solar cells from 14 different deposition runs are evaluated. From each sample, 10 solar cells are manufactured. For the analysis of the current-voltage characteristics of the solar cells, only those solar cells are taken into account, where efficiencies deviate less than 1 % absolute from the highest efficiency for each sample.

- AIST: The CIGS absorber layer is grown by a static multi-stage co-evaporation process. At the end of the process a KF PDT is carried out *in-situ*, and the solar cells are finished using a CdS buffer layer, a sputtered ZnO/Al:ZnO window layer, and Ni/Al contact grid [27]. The efficiency of a solar cell yields an efficiency of 21.6 % (with anti-reflective coating). The PL peak position is around 1.13 eV.
- Empa: The CIGS absorber layer is grown by a static multi-stage co-evaporation process, where Ga is supplied only during the first stage [28]. A RbF PDT is carried out in-situ and the solar cell yields an efficiency of 19.2 % (with anti-reflective coating), which is the same device as published elsewhere [29]. The Ga is located only towards the back contact for passivation purposes [28] and the PL peak position is at 1.0 eV, which corresponds to a $CuInSe_2$ stoichiometry with GGI = 0.

Experimentally, the differential diode factor is measured by intensity-dependent PL spectroscopy using eqn. (1). Illumination is provided by a 660 nm wavelength laser. The intensity of the laser beam is varied by the output power of the laser as well as by optical density (OD) filters. The generation flux is calibrated at the sample position, from where the PL light is collected. First, a power meter measures the power of the total beam area and second a camera captures the beam shape to determine the spot size (approximately 1.1 mm radius). Subsequently, the power is calculated in the center of the Gaussian spot, from where the PL is collected.

Polycrystalline CIGS absorbers have a sufficiently high PL yield, which allows the reduction of the generation flux by several orders of magnitude below 1-sun excitation. 1-sun excitation corresponds to the same absorbed photon flux as realized by illumination with an AM1.5G spectrum. Thus, it is reasonable to evaluate the differential diode factor according to eqn. (1). To reduce the noise in the diode factor when calculating the derivative, it proved to be important to reduce statistical errors for the determination of the generation flux as much as possible. Therefore, the calibration mentioned above (power and beam spot size measurement) is carried out for each setting of laser output power and OD filter. Additionally, a CdS buffer layer is deposited on the front surface for passivation purposes and to prevent degradation during PL measurements [30, 31]. For the samples without the addition of Ag, i.e. the samples from Empa, AIST, and ZSW without Ag, the processing after the CdS layer is omitted, i.e. no transparent conductive oxide was deposited. For the sample including Ag during the deposition, the absorber was only available in the form of a finished solar cell device. For the PL study, the absorber was then etched in acetic acid (AcOH), which is known to remove the window layers but not the CdS buffer layer [32]. Here, the sample was etched in 5 % AcOH for 1 minute in an ultrasonic bath. The removal of ZnO and the presence of CdS after the etching is checked by energy dispersive X-ray spectroscopy (Supplementary Figure 3).

The generation flux dependence of the diode factor for the polycrystalline CIGS samples is fitted with the model of metastable defects as described in section 2. Additionally, boundary conditions forcing a free hole density $p_0$ in the dark (i.e. the lowest generation fluxes) greater than $p_{0,bound}$ was applied to several fits. The boundary conditions was implemented in the fitting routine as a 'soft' bound so that the residuals are scaled by a factor $f_{bound}$ defined as

$$f_{bound} = \begin{cases} 1 + \log_{10}(p_0) - \log_{10}(p_{0,bound}) \ for\ p_0 > p_{0,bound} \\ 1\ otherwise \end{cases}$$

Thus, the following expression is minimized

$$\mathrm{X}^2 = \frac{1}{N}\sum_i \left[f_{bound}\left(A^i_{exp} - A^i_{calc}\right)\right]^2$$

where $A^i_{exp}$ and $A^i_{calc}$ are the experimental and calculated diode factors and $N$ denotes the number of data points (number of experimentally determined diode factors $A^i_{exp}$). Errors are calculated from the covariance matrix of the fit result.

Epitaxial films have a lower PL yield and therefore the intensity could be varied only by roughly one order of magnitude (from ~1/2 to ~4 sun excitation). In that case, the diode factor is extracted from a linear fit. As will be shown below, the diode factor is only slowly varying with generation flux. Thus, the linear fit gives a good estimate for this range of generation fluxes. Measurements are carried out without a CdS buffer layer. Instead, the samples are measured in a $N_2$ atmosphere to prevent degradation.

Raman spectra were recorded at room temperature with a 532 nm laser using a 50x objective lens and a numerical aperture of 0.5 in combination with a 2400 lines/mm grating. An edge/notch filter is used to block the incident laser beam.

Current density versus voltage (J-V) characteristics are measured in the respective institutes of device fabrication under standard test conditions. A fitting routine based on a 1-diode model (eqn. (7)) is used for the illuminated J-V characteristics to extract diode parameters.

$$J(V) = J_0\left[\exp\left(\frac{q(V - r_s J)}{A_{el} k_B T}\right) - 1\right] + \frac{V - r_s J}{R_{sh}} + J_{ph} \qquad (7)$$

Where $J_0$ is the saturation current density, $A_{el}$ the electrical diode factor, $r_s$ the series resistance, $R_{sh}$ the shunt resistance and $J_{ph}$ the photo-current. In the 1-diode model, in general, the photo current is determined as $J_{ph} = J_{sc}(1 + r_s/R_{sh})$, where $J_{sc} = J(0)$ is the short-circuit current density. For the solar cells investigated here it holds that $r_s \ll R_{sh}$ and thus $J_{ph} \approx J_{sc}$. The 1-diode fit is subsequently carried out for the data $J(V) - J_{sc}$, which proved to be much more robust than having $J_{ph}$ as an additional fit parameter. Two fitting routines are used: i) the iv-fit routine [33] and ii) a self-implemented fitting routing using an orthogonal distance regression with logarithmic values for the current density. Error bars $\pm e_i$ of the fit parameters $p_i$ are computed as $e_i = \left(p_i^{i)} - p_i^{ii)}\right)/2$, where $p_i^{i)}$ and $p_i^{ii)}$ are the fit parameters of method i) and ii), respectively.

The fill factor (FF) of a J-V curve is given by

$$FF = \frac{V_{mpp}J_{mpp}}{V_{oc}J_{sc}} \tag{8}$$

where $V_{oc}$ is the open-circuit voltage and $V_{mpp}$ and $J_{mpp}$ are the voltage and current density at the maximum power point (mpp). In the 1-diode model (eqn. (7)) the FF depends mainly on $V_{oc}$ (Supplementary Figure 4). To compare the FF for solar cells with different $V_{OC}$ values, the FF is extracted from a calculated J-V curve using the fit parameter, whereas $J_0$ is adjusted according to

$$J_{0,fix} = \frac{-J_{sc} - \frac{V_{OC,fix}}{R_{sh}}}{\exp\left(\frac{qV_{OC,fix}}{A_{el}k_BT}\right)} \tag{9}$$

With eqn. (9), $J_0$ is scaled such that the J-V curve yields an open-circuit voltage of $V_{OC,fix}$. Subsequently, $FF_{fix}$ is extracted from the J-V curve calculated using $J_{0,fix}$. It is noted that $FF_{fix} \neq FF$. However, this procedure allows the comparison of fill factors from J-V curves with different $V_{OC}$ values and takes into account non-negligible and individual contributions of parasitic resistances, which is not possible with previous empirical expressions [34].

# 4. Results

## 4.1. Epitaxial CIGS absorbers

For p-type CIGS, $A > 1$ is due to a downshift of the hole quasi Fermi level in low-injection conditions. In the model of metastable defects [7] (see also section 2), electron injection converts metastable donors into acceptors. In contrast, for n-type CIGS films, electrons are majority carriers and consequently exist abundantly compared to p-type CIGS. Within the same model, therefore, all metastable defects exist in the acceptor configuration and do not change their state to a donor configuration upon excitation of the n-type CIGS layer. The reason is the dominating electron capture process ($\tau_{EC}^{-1}$) due to the large number of free electrons and the small contribution from the hole capture process ($\tau_{HC}^{-1}$) due to the scarcity of holes. Thus, a diode factor of 1 is expected.

Figure 2a shows the spectrally integrated PL yield as a function of the generation flux (linear to laser power) for the epitaxial absorber layers. N-type CIGS samples with a $GGI \leq 0.15$ are shown in blue, while p-type samples with $GGI \geq 0.19$ are plotted in orange. Except the sample with the highest PL yield, it is visible by eye that the p-type samples have a higher slope and therefore a higher diode factor (eqn. (1)) compared to the n-type samples. Linear fits are performed (dashed lines) and the diode factor (slope of the fit) is plotted in Figure 2b versus the GGI. Clearly, a jump in the diode factor is observed at the transition between n-type and p-type CIGS. Also, n-type films have $A = 1$ as expected for SRH recombination. A single n-type absorber exists having A > 1. This particular sample exhibits a much broader PL peak than the other samples, has the lowest CGI as well as an increased contribution of an ordered defect compound seen from Raman spectroscopy [35] (Supplementary Figure 6). We assume that in this sample the PL signal originates not from a single-phase n-type absorber, which is the reason for the increased PL yield as well as the increased diode factor.

The results on epitaxial CIGS films are a strong indication for the model of metastable defects responsible to cause diode factors greater than one in p-type CIGS. In particular, it demonstrates that it is a bulk effect and not caused by grain boundaries or alkali elements. Still, alkali elements are known to change the net doping of the absorber layer and with that the hole quasi Fermi level in equilibrium. Thus, an indirect influence of alkalis on the diode factor may be anticipated.

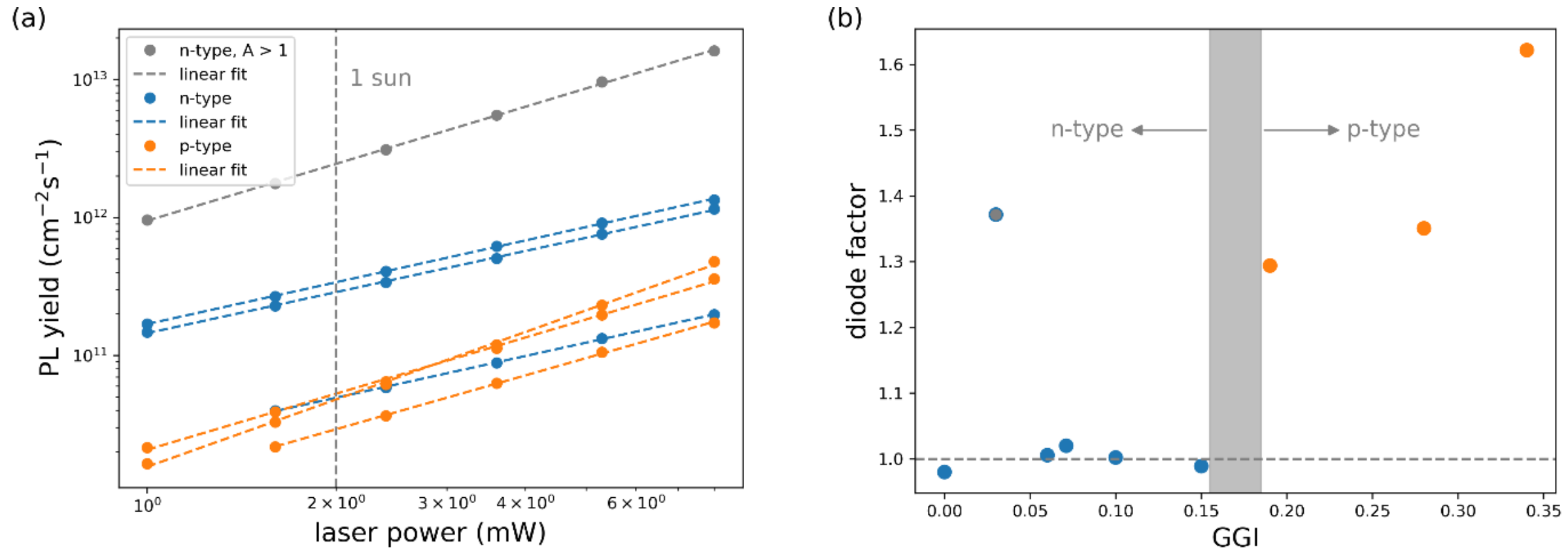

Figure 2 – a) measured and spectrally integrated PL yield for epitaxial CIGS absorbers with different GGI values. A linear fit is applied to determine the diode factor from the slope. b) Diode factor versus the GGI. A transition from $A = 1$ to $A > 1$ is observed, where the semiconductor turns from n-type to p-type [16].

## 4.2. Polycrystalline CIGS absorbers

Four polycrystalline CIGS absorbers are investigated for the generation flux dependence of the diode factor. The J-V characteristics for the best solar cells fabricated from the same absorber are shown in Supplementary Figure 5. The solar cell parameters are summarized in Table 3. The generation flux dependence of the spectrally integrated PL yield for the polycrystalline CIGS samples is shown in Figure 3a. The differential diode factors (eqn. (1)) are plotted in Figure 3b (solid symbols, dashed lines are added as a guide to the eye). Both samples with a PL peak position around 1.1 eV (inset of Figure 3a) and without the addition of Ag, i.e. emission from a $CuIn_{1-x}Ga_xSe_2$ with x ≈ 0.15 (light blue and orange dots), have a diode factor of approximately 1.3 at 1-sun equivalent excitation (gray vertical bar) and an increasing diode factor with decreasing generation flux. The sample with Ga located only towards the back contact (also without Ag) and a PL peak emission centered around 1.0 eV shows a rather constant diode factor of 1.4 (red dots). Interestingly, with the addition of Ag, i.e. a $(Ag,Cu)(In,Ga)Se_2$ absorber, the diode factor drops significantly to a value of approximately 1.1.

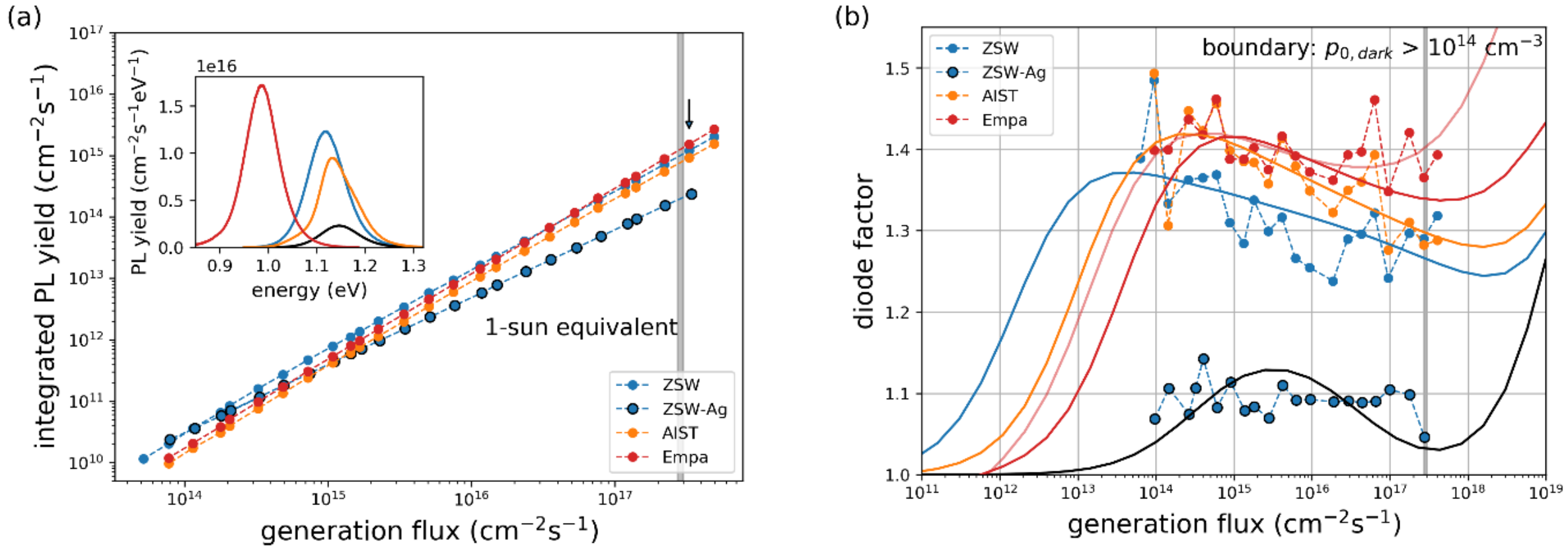

Figure 3 – a) Experimentally measured spectrally integrated PL yield as a function of the generation flux for four polycrystalline $Cu(In,Ga)Se_2$ absorber layers. Gray vertical bar indicates a 1-sun equivalent generation flux. Inset shows measured PL spectra at equivalent generation fluxes just above 1-sun (black arrow). b) Calculated differential diode factor according to eqn. (1) (full symbols). Solid (dark) lines represent best fits following the model of metastable defects with the boundary condition of a minimum free hole density of $10^{14}$ $cm^{-3}$. Data from the Empa sample is the only case, where a significant better fit is obtained by loosening the boundary condition to $p_{0,bound} = 10^{13}\ cm^{-3}$.

The model of metastable defects is employed to fit the experimental data. The result with the boundary condition of a minimum free hole density in the dark of $10^{14}$ $cm^{-3}$ is shown as solid dark lines in Figure 3b. The parameters for the fit are listed in Table 2. Using the fitting parameters, the diode factor is calculated additionally outside the experimental range of generation fluxes. The trend of the generation flux dependence and the magnitude of the diode factor are well described by the fits. In particular, for the ZSW, AIST, and Empa samples, the increased diode factors between 1.3 and 1.4 and the generation flux dependence is reproduced well. As elaborated in section 2 (Theoretical background), a generation flux independent diode factor of 1.1 over four orders of magnitude is more difficult to describe. Yet, this rather simple model can describe the observed diode factors between 1.4 and 1.1. It is interesting to note that the activation energies for the two CIGS samples (ZSW and AIST) are the same within error. $\Delta E_{HC}$ for the Ag-containing sample is higher, indicating that the presence of Ag in the crystal changes the dynamics of this metastable defect. For the Empa sample, a considerable better fit is obtained when the boundary condition is relaxed to $p_{0,bound} = 10^{13} cm^{-3}$, which is depicted in Figure 3b as the light red curve. In particular, the improved fit with $p_{0,bound} = 10^{13} cm^{-3}$ results in considerably lower $\Delta E_{HC}$ and $\Delta E_{EE}$ values for the $CuInSe_2$ sample (Empa), indicating an even stronger influence of the Ga content.

This observation is not unexpected, since the dynamics of the defect depends on the dynamics of the group-III dimer near the Se vacancy [36]. An overview of the fit quality ($\mathrm{X}^2$ values) and the energetic barriers $\Delta E_{HC}$ and $\Delta E_{EE}$ with the dependence on the fit boundary condition can be found in Supplementary Figure 10. Supplementary Figure 11 shows the fits for the various boundary conditions visually together with the free hole density as a function of the generation flux for all four samples.

Table 3 – Solar cell parameters for the four polycrystalline CIGS devices investigated by PL spectroscopy (Figure 3).

| Sample | Efficiency (%) | Fill factor (%) | $V_{OC}$ (V) | $J_{SC}$ (mAcm$^{-2}$) |
|---|---|---|---|---|
| ZSW | 18.8 | 78.1 | 0.737 | 32.7 |
| ZSW-Ag | 18.1 | 81.0 | 0.703 | 31.8 |
| AIST [a] | 21.6 | 80.2 | 0.778 | 34.6 |
| Empa [a] | 19.2 | 74.5 | 0.609 | 42.3 |

[a] with anti-reflective coating

## 5. Discussion

CIGS as used in solar cells is a p-type semiconductor with net p-type doping levels around $10^{15}$ to $10^{16}$ cm$^{-3}$ [9, 10]. Using a carrier lifetime of $\tau = 100$ ns, an absorber thickness $d = 2\ \mu m$ and eqn. (4), the excess carrier density $\Delta n$ is estimated to be $10^{14}$ cm$^{-3}$ at 1-sun equivalent generation flux and as low as $5 \times 10^{10}$ for a generation flux of $10^{14}$ cm$^{-2}$s$^{-1}$ (see also Supplementary Figure 8). Clearly, the absorber is in low injection for the experimental conditions of the PL measurements. Thus, a diode factor of 1 is expected for recombination in the quasi neutral region, i.e. in the absence of a pn-junction [37]. The fact that the model of metastable defects [7] can describe the diode factor of $A \approx 1.4$ for generation fluxes as low as $10^{14}$ cm$^{-2}$s$^{-1}$ is a strong hint for the validity of the model.

The parameters for the fits shown in Figure 3b dictate that the free hole density in the dark, i.e. sufficiently low generation fluxes, is rather low between $10^{13}$ and $10^{14}$ cm$^{-3}$. This finding is in contradiction to results obtained from other techniques such as capacitance-voltage profiling [10, 38], electron beam induced current [39] or Hall [9, 40, 41] measurements, which report values around $10^{15}$ and $10^{16}$ cm$^{-3}$ [10]. Supplementary Figure 11 shows the fit to the experimental diode factors and the free hole density with various boundary conditions up to $p_{0,bound} = 10^{15} cm^{-3}$. Even though the quality of the fit deteriorates, it is still interesting to note that a qualitative agreement is found. At this point it is also stressed that only a very simply model is used here, i.e. a homogenous absorber (without bandgap grading), a single SRH recombination channel in the bulk of the absorber, no surface recombination, and only a single metastable defect.

In refs. [42, 43], the authors present a voltage, i.e. generation flux, independent diode factor greater than 1 due to exponential defect distributions from the band edges into the bandgap region. However, these defect distributions only yield a diode factor greater than 1 for recombination where $n = p$, i.e. in the space charge region, which is not the case for the material under investigation here. In particular, for a doped semiconductor under low-injection conditions and recombination in the quasi neutral zone, as in a PL measurement, any defect distribution will result in a diode factor equal to one [37].

Based on the discussion one can ask if other factors than metastable defects may contribute to the increased diode factor (in low injection conditions). A possible scenario might be the presence of space-charge-regions (SCR) within the absorber, possibly caused by n-type doping of the CdS buffer layer [44] or charges located at grain boundaries. These SCR could then result in regions where $n \approx p$. The diode factor is then influenced by exponential defect distributions (see above), i.e. band tailing, and high injection conditions in these regions. However, we disregard this explanation due to the following points:

- A small SCR results only in a small band bending of the conduction and valence band. Thus, in low injection conditions, here for generation fluxes $< 10^{-3}$ equivalent suns, electrons are still minority carriers. Thus, even though SCRs exist, the absorber layer is in low-injection conditions everywhere and a diode factor of 1 is expected for SRH recombination.
- Large SCRs would result in charge carrier separation effects. These effects can be observed well by time-resolved PL measurements and is evidenced by strongly reduced decay times of the transient [45]. However, CIGS absorber layers covered only with a CdS buffer layer generally do not exhibit such charge separation effects [45]. For instance, $CuIn(Ga)Se_2$ samples with a Ga grading only towards the back contact, similar to the Empa sample investigated here, show lifetimes as high as 400 ns [29].
- The increased diode factor is also observed for p-type epitaxial absorber layers (section 4.1). These absorber layers are measured without a CdS buffer layer and do not incorporate grain boundaries.

It is interesting to note that the activation energy for the hole capture process $\Delta E_{HC}$ is fitted to be higher for the Ag-containing CIGS sample (ZSW-Ag) compared to the Ag-free samples with similar bandgaps (ZSW and AIST) as listed in Table 2 (see also Supplementary Figure 10b). The higher activation energy $\Delta E_{HC}$ is evidenced complementarily by admittance spectroscopy measured on finished solar cell devices. Supplementary Figure 12 shows the measured capacitance as a function of the temperature in the relaxed and light-soaked state (see ref. [7] for measurement details). The relaxation process happens for the Ag-containing CIGS sample at higher temperatures, which indicates a slower time constant and thus a higher activation energy $\Delta E_{HC}$ for the hole capture process. It is stressed here that the values for the activation energies are not absolute as the activation energy for the electron capture process $\Delta E_{EC}$ is fixed to 0.35 eV for the fitting routine. In order to obtain absolute values, dynamic measurements need to be carried out. Nevertheless, the activation energies can be compared relatively to each other. It is found here that the activation energy for the hole capture process $\Delta E_{HC}$ is smaller than the activation energy of the electron capture process $\Delta E_{EC}$. This trend is in contrast, i.e. $\Delta E_{EC} < \Delta E_{HC}$, to values predicted theoretically [11] and used experimentally to model the doping profiles after voltage bias soaking from capacitance-voltage measurements [13]. To get a better understanding of the absolute values of the activation energies, dynamic measurements will be needed in combination with solutions of the time dependent occupancy function $f_A$.

In order to improve device efficiency, metastable defects responsible for the increased diode factors need to be avoided. Illuminated J-V curves from a large set of samples from ZSW, AIST and Empa are evaluated and the electrical diode factor is extracted from fitting (see experimental section). Here, the electrical diode factor is distinguished from the diode factor measured optically and introduced above (from now on labelled as optical diode factor). The reason is that upon finishing the solar cell, additional recombination channels might be introduced, which can lead to deviations of the electrical diode factor from the optical diode factor [8]. To reduce scattering of the data for the FF, $J_0$ of the J-V curves is scaled such that every curve yields $V_{OC,fix} = 0.7$ V (see experimental section). Figure 4 clearly shows that the fill factor increases as the electrical diode factor decreases. In particular, assuming realistic values for the series and shunt resistance under illumination (Supplementary Figure 13), the diode factor is the main contributor to the fill factor (gray solid line) according to the 1-diode model (eqn. (7)). The simulated fill factors demonstrate that the FF can be increased by 1 % absolute if losses due to parasitic series or shunt resistances can be eliminated, respectively, or by 2 % if both are eliminated. Vertical dashed lines indicate the optical diode factors (Figure 3b) averaged over all generation fluxes. The electrical diode factors from the exact same pieces of absorbers are shown as the highlighted symbols (labeled with ‘PL study’). It is apparent that the optical diode factors serve as a lower bound for the electrical diode factors. This behavior is expected when assuming additional space-charge region recombination taking place in a finished solar cell device, as space-charge region recombination has a diode factor close to two [37]. In particular, the best solar cells (here from AIST), have electrical diode factors very close to the optical diode factor, while the deviations become larger for worse performing devices.

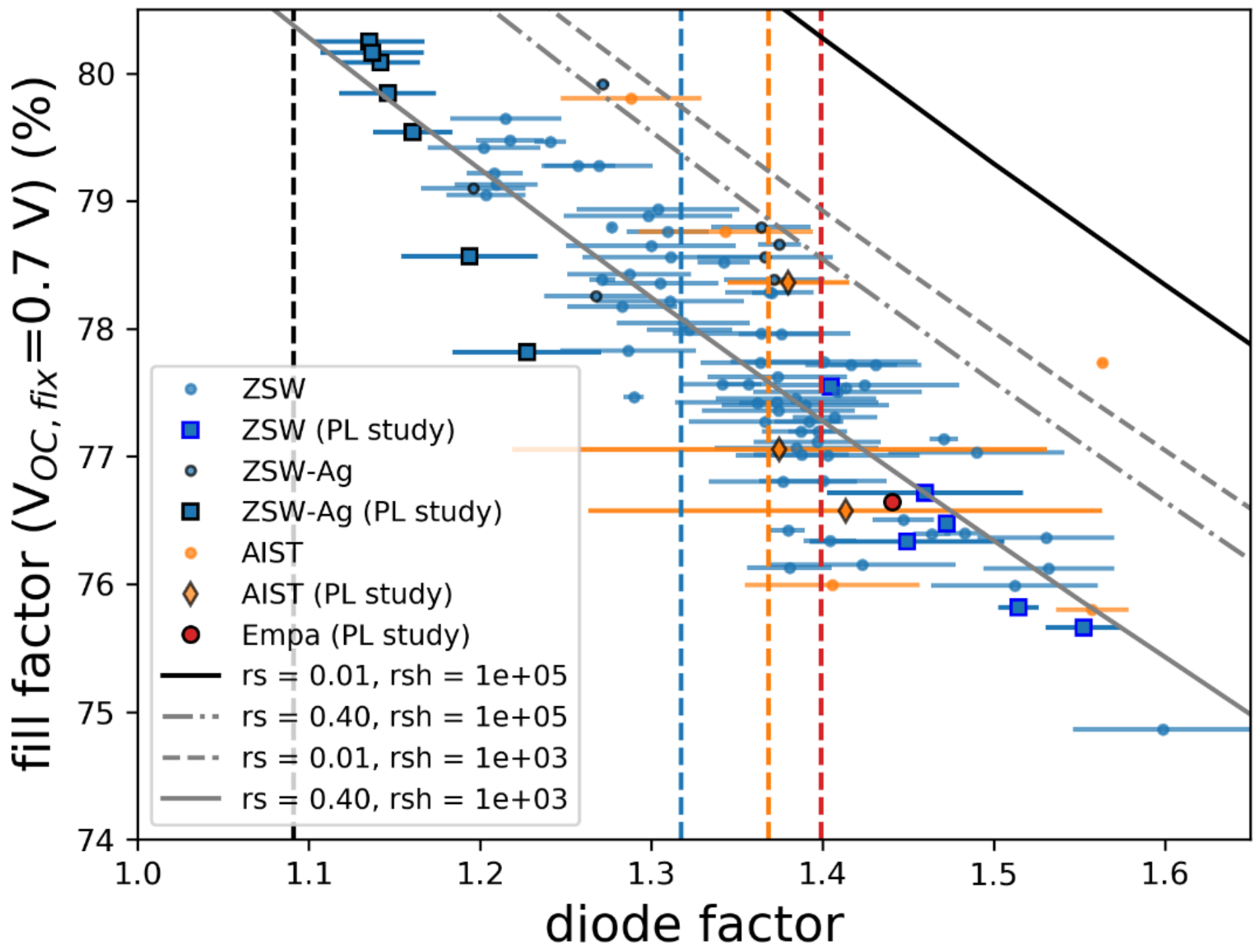

Figure 4 – Fill factors of finished solar cell devices as a function of the electrical diode factor (symbols). Highlighted symbols correspond to solar cells from the same absorber as used for the measurement of the optical diode factor by PL spectroscopy. Dashed vertical lines show the mean optical diode factors computed by averaging over the generation flux dependent data (see also Figure 3b). Gray and black lines show simulated FF based on the 1-diode model (eqn. (7)) with (gray lines) and without (black line) parasitic resistances (units of $\Omega cm^2$).

Noteworthy, the addition of Ag during the CIGS deposition results in the lowest diode factors and consequently in the highest fill factors of 81.0 % at a $V_{OC}$ of 703 mV for the best device (shown in Supplementary Figure 5). With the addition of Ag, Essig et al. reported a 20.5 % efficient device (with anti-reflective coating) with a FF of 80.8 % at a $V_{OC}$ of 745 mV, which also represents a good fill factor [25]. It is noted that this sample was prepared in the same deposition tool as the samples investigated here from ZSW. At the moment, it is not clear how Ag influences the electrical properties of the CIGS material and how the metastable defect density is influenced. In reference [46] (see Table 1 therein) the addition of Ag to CIGS also seems to improve the FF. However, the reported record device (20.9 % efficiency) has a diode factor of 1.4 and only a FF of 76.8 % at a $V_{OC}$ of 814 mV. In contrast, in 2008 a CIGS solar cell prepared at NREL obtained a FF of 81.2 % at a $V_{OC}$ of only 690 mV, which was fabricated without Ag and without alkali PDT [47]. Indeed, this particular sample had the lowest diode factor of 1.14 among the studied samples. Interestingly, one of the steps for improvements is described by an in-situ high temperature annealing at substrate growth temperature only in Se atmosphere. Such an annealing could have an influence on the density of $V_{Cu}$-$V_{Se}$ metastable defect complexes. Thus, at the moment it is not clear if Ag influences the metastable defect density directly or indirectly. Nevertheless, to obtain higher FF values, the diode factor needs to be reduced with the help of decreasing the influence of metastable defects.

# 6. Conclusion

The consequences of the $V_{Cu}$-$V_{Se}$ metastable defect on the diode factor in Cu(In,Ga)$Se_2$ absorber layers and solar cells is investigated. Due to the fact that the $V_{Cu}$-$V_{Se}$ changes from a donor to an acceptor state upon electron injection, the hole quasi Fermi level moves towards the valence band, which results in diode factors greater than one, even in low injection conditions. Using this model we can explain

i) that n-type epitaxial Cu(In,Ga)$Se_2$ absorbers have a diode factor of one, because electrons are majorities. Under these circumstances, all metastable defects exist in the acceptor configuration and do not change their charge state upon injection of holes.

ii) that p-type epitaxial Cu(In,Ga)$Se_2$ absorbers have a diode factor greater than one as expected in the presence of metastable defects. These results show that it is indeed an intrinsic bulk property of the Cu(In,Ga)$Se_2$ alloy.

iii) the generation flux dependence (i.e. density of excess minorities) of the diode factor for high quality polycrystalline Cu(In,Ga)$Se_2$ films. Experimentally, a small dependence of the diode factor with respect to the generation flux of the laser was measured, which can well be described by the model of the $V_{Cu}$-$V_{Se}$ defects.

In addition, we have evaluated a large set of solar cells and their corresponding electrical diode factors from J-V characteristics. We find that the diode factor measured optically on the bare absorber imposes a lower limit for the electrical diode factor as measured on complete solar cells. In addition, a clear trend is observed that the diode factor limits the fill factor. Interestingly, absorber layers alloyed with Ag, i.e. (Ag,Cu)(In,Ga)$Se_2$, demonstrate the lowest diode factors, which in turn lead to the highest fill factors of 81.0 %. Thus, a better understanding of the impact of Ag on the electrical properties of Cu(In,Ga)$Se_2$ is required, in particular, its effect on metastable defect transitions. We have demonstrated that metastable defects increase the diode factor and thus reduce the fill factor and the efficiency and that their dynamics depends on the composition of the absorber.

# 7. Acknowledgements

This work was supported by the Luxembourgish Fond National de la Recherche in the framework of the project C17/MS/11696002 GRISC and by the German Federal Ministry for Economic Affairs and Climate Action (BMWK) within the EFFCIS-II and ODINCIGS projects under contract numbers 03EE1059A and 03EE1078, respectively. The authors thank Ricardo Poeira for his assistance with the Raman measurements.

Table 4 – Parameters for the simulation of the diode factor. The free electron mass is denoted as $m_e$.

| **Parameters** | **Description** | **Value** |
|---|---|---|
| | | |
| T / K | Temperature | 296 |
| $E_g$ / eV | Bandgap | 1.10 |
| $N_A$ / cm$^{-3}$ | Doping density | $10^{16}$ |
| $m_e^*$ | Electron effective mass | 0.1 $m_e$ |
| $m_h^*$ | Hole effective mass | 0.9 $m_h$ |
| B / cm$^3$s$^{-1}$ | Radiative recombination constant | $1.28 \cdot 10^{-10}$ |
| | | |
| $\tau_{SRH}$ / ns | Non-radiative lifetime for electrons and holes | 100 |
| $E_{t,SRH}$ / eV | Distance of defect from the Valence band | 0.6 |
| | | |
| $\Delta E_{EC}$ / eV | Barrier for electron capture process | 0.35 |
| $\Delta E_{EE}$ / eV | Barrier for electron emission process | Fit-parameter |
| $\Delta E_{HC}$ / eV | Barrier for hole capture process | Fit-parameter |
| $\Delta E_{HE}$ / eV | Barrier for hole emission process | Calculated (eqn. (5) and (6)) |
| $N_{MS}$ / cm$^{-3}$ | Density of metastable defects | Fit-parameter |

# Supplementary Information

## A. Additional Discussion on the rate constants $\tau_{ij}^{-1}$

The rate constants $\tau_{ij}^{-1}$ as a function of the generation flux are shown in Supplementary Figure 3 for the cases where the parameters of the metastable defect are fitted to yield diode factors of A=1.2, A=1.3, and A=1.4.

**$A = 1.3$**

The rate constants $\tau_{EC}^{-1}$, $\tau_{EE}^{-1}$ and $\tau_{HC}^{-1}$ vary as a power law (linear on log-log scale) over a wide range of generation fluxes. The steadily increasing contribution of $\tau_{EC}^{-1} \propto n \propto G$ leads to a conversion of metastable defects into the acceptor state. Due to that increase in the net doping density, also $\tau_{HC}^{-1}$ and $\tau_{EE}^{-1}$ increase until they level off between generation fluxes of $10^{17}$ and $10^{18}$ cm$^{-2}$s$^{-1}$, where $f_A$ converges to one. Thus, the diode factor is increased over a broad range of generation fluxes. For generation fluxes greater than $10^{19}$ cm$^{-2}$s$^{-1}$, the system is in high-injection conditions resulting in a stronger increase of $\tau_{HC}^{-1}$ and $\tau_{EE}^{-1}$ as photo-generated holes significantly contribute to the hole density in the valence band.

**$A = 1.2$**

Compared to the case of A=1.3, the free hole density for the lowest generation flux is higher (see Supplementary Figure 8). Consequently, the hole quasi Fermi level is closer to the valence band. Thus, the conversion of the metastable defects from donor to acceptor result in a smaller shift of the hole quasi Fermi level and therefore in a smaller diode factor. For the lowest generation fluxes, the hole capture process (time constant $\tau_{HC}^{-1}$) dominates also in the case of A=1.2. Thus, to have the transition from the metastable donor to the acceptor in the same range of generation fluxes between $10^{13}$ and $10^{17}$ cm$^{-2}$s$^{-1}$, the activation energy for the hole capture needs to be higher (compare Table 2). For generation fluxes $\gtrsim 10^{17}$ cm$^{-2}$s$^{-1}$, almost all metastable defects are in the acceptor configurations and the diode factor decreased again towards 1.

**$A = 1.4$**

Compared to the case of A=1.3, the free hole density for the lowest generation fluxes is lower (see Supplementary Figure 8), which allows for a greater shift of the hole quasi Fermi level and thus a larger diode factor. With increasing generation flux, the electron density increases, which results in small changes of the occupation of the metastable defect in the acceptor state ($f_A$ in Supplementary Figure 3). Due to the low hole density, already a small change in $f_A$ results in a considerable shift of the hole quasi Fermi level, resulting in a diode factor of approximately 1.4. It is interesting to note that in this case, $f_A$ never approaches 1. For generation fluxes $\gtrsim 10^{19}$ cm$^{-2}$s$^{-1}$, the semiconductor is in high injection and therefore the occupancy of $f_A$ shifts towards 0 (see explanation for the case of A=1.3).

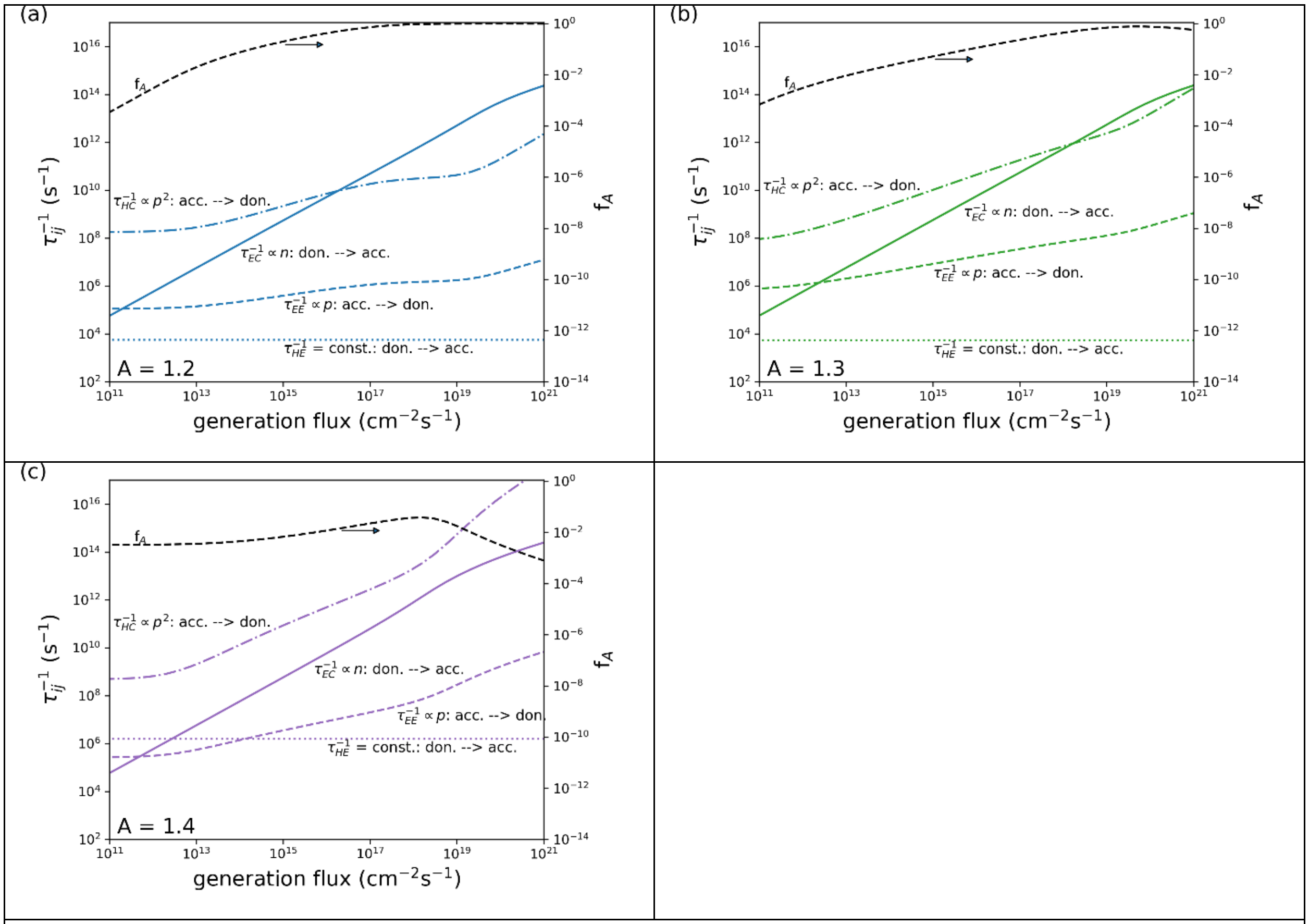

Supplementary Figure 1 - Rate constants $\tau_{ij}^{-1}$ for the four different transition process (Table 1) of the metastable defects between donor and acceptor state for the fits of an approximately constant diode factor of 1.2 (a), 1.3 (b), and 1.4 (c) see Fig. 1(c).

The simulations presented above show that the model of metastable defects (eqn. (3)) can describe very well a generation flux independent diode factor of 1.30 – 1.35. For diode factors outside this range, small deviations are expected such as concave or convex shapes with respect to the generation flux.

## B. Supplementary Figures

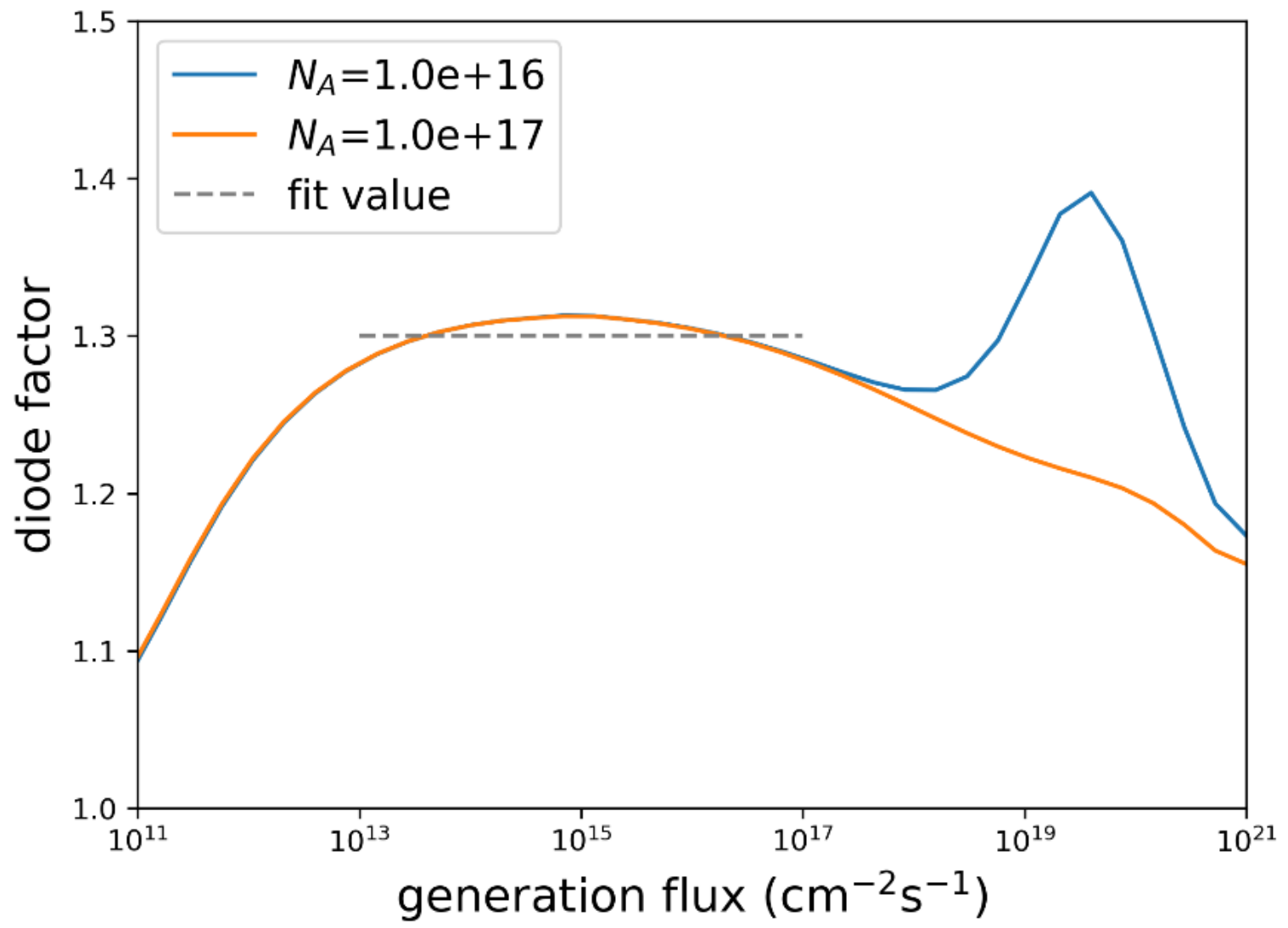

Supplementary Figure 2 – Comparison of the fitted diode factors (solid lines) using different doping densities. As described in section 2, the doping density is a fixed parameter, while the metastable defect density is allowed to vary. The main difference occurs outside the experimental accessible range of generation fluxes (otherwise heating of the sample needs to be taken into account) at generation fluxes greater than $10^{18}$ $cm^{-2}s^{-1}$. The peak in the diode factor for $N_A = 10^{16}$ $cm^{-2}s^{-1}$ is due to high injection and dominating SRH recombination. The curve with $N_A = 10^{17}$ $cm^{-2}s^{-1}$ does not show this peak (or only as a shoulder at roughly 1 order of magnitude higher generation fluxes) because radiative recombination is already the dominating recombination channel.

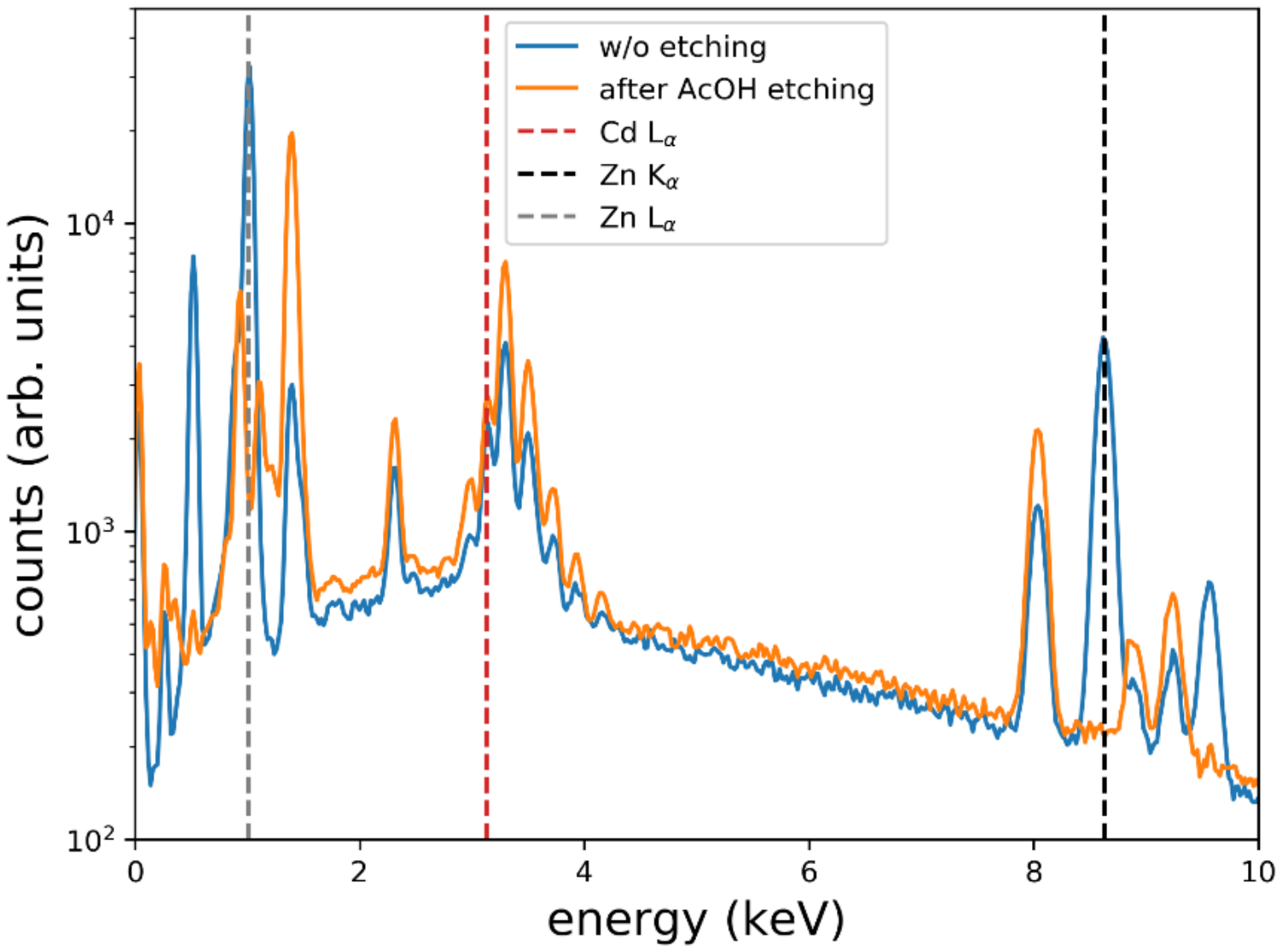

Supplementary Figure 3 – Energy dispersive X-ray (EDX) spectra for the Ag-containing solar cell prepared at ZSW before (blue) and after AcOH (orange) etching. As indicated by the gray and black dashed lines, the Zn signal is reduced below the detection limit, indicating that the AcOH etching removes the window layers with the Ni/Al grid fingers on top. However, the thin CdS buffer layer remains at the absorber surface as the Cd signal (red dashed line) is still detectable without any significant reduction in intensity.

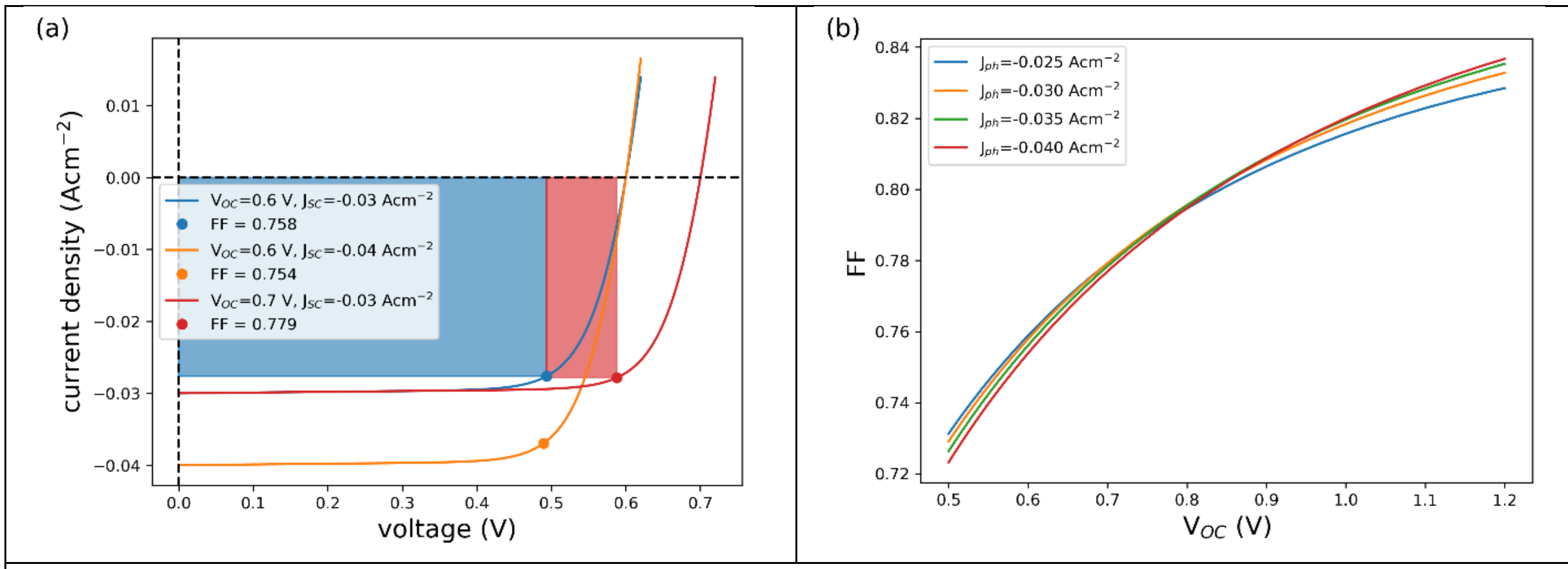

Supplementary Figure 4 – a) Simulated J-V characteristics with fixed series and shunt resistances of 0.5 and 1000 $\Omega cm^2$ and a diode factor of 1.3. Solid circles indicate the maximum power point. The fill factor increases with $V_{OC}$ as the portion of the rectangle takes a bigger contribution of the total area enclosed by the IV curve in the 4th quadrant. b) Calculated fill factors with respect to $V_{OC}$ for different short-circuit current densities. Clearly, $J_{SC}$ only has a minor effect on the fill factor.

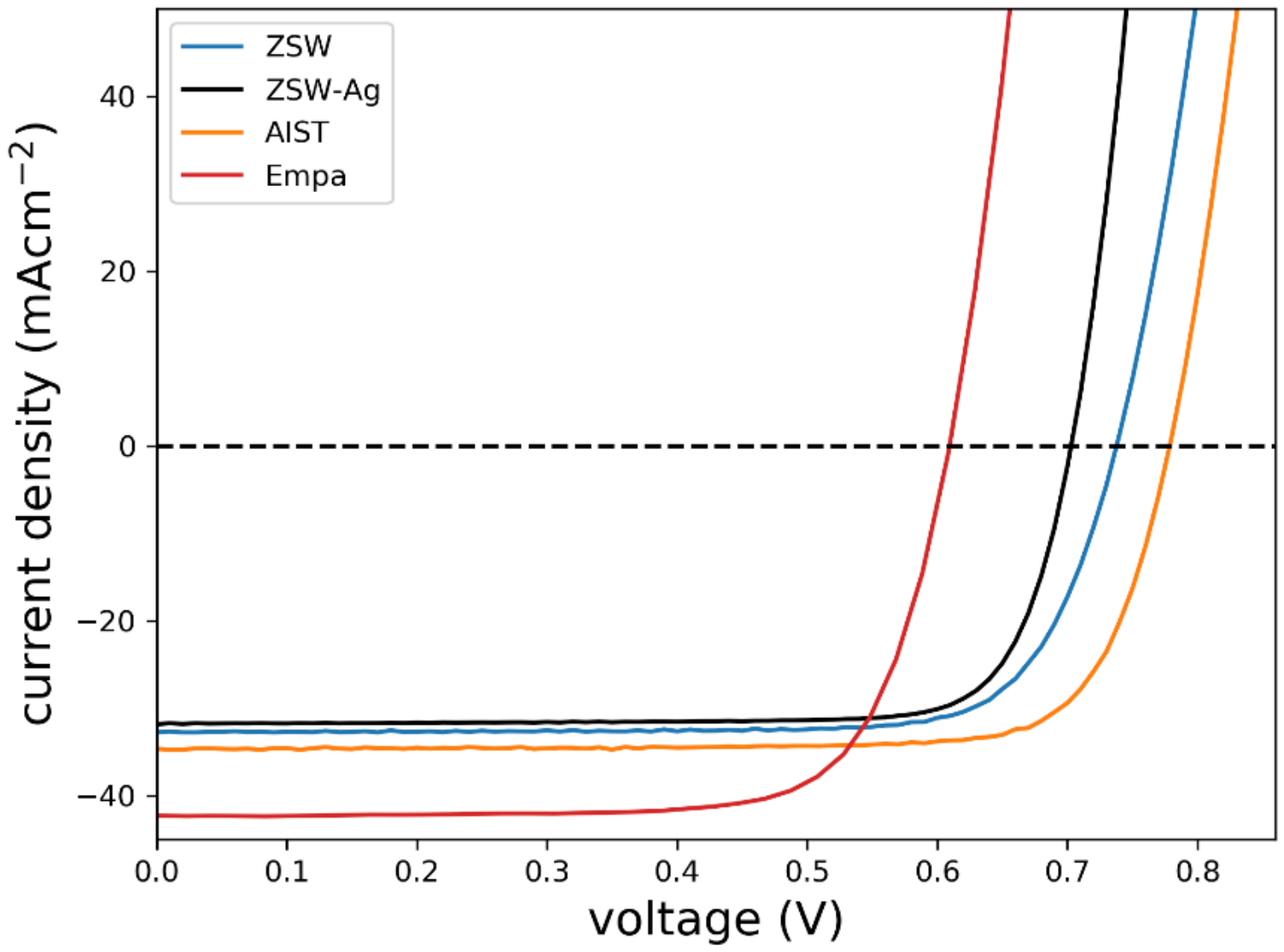

Supplementary Figure 5 – Best J-V curves from the samples, which are used for the study of the diode factor by photoluminescence spectroscopy. The sample from AIST and Empa have an anti-reflection coating. The solar cell parameters are summarized in Table 3.

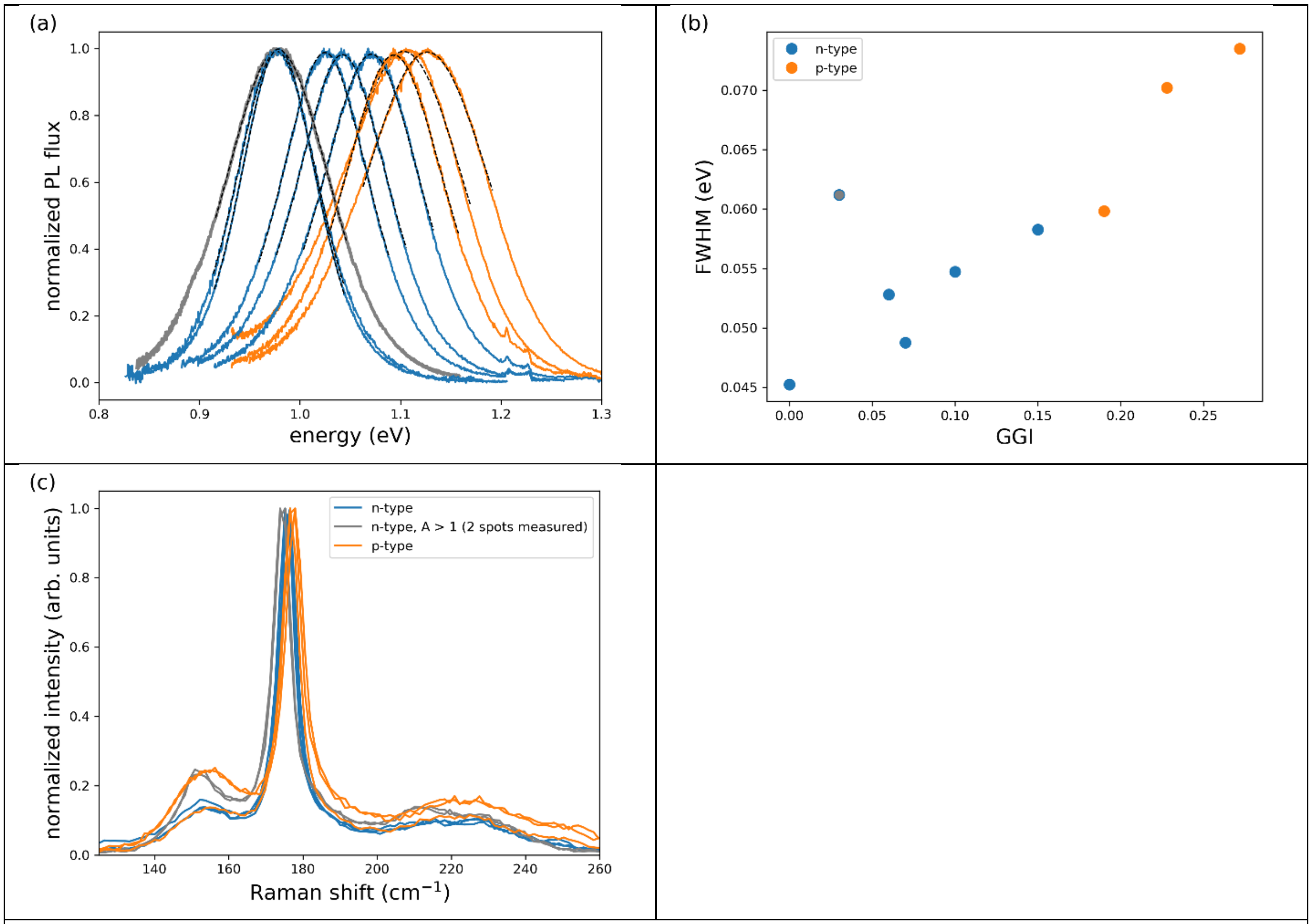

Supplementary Figure 6 – a) normalized PL spectra for the epitaxial CIGS absorber layers presented in Figure 2. Black dashed line is a fit using a Pseudo-Voigt function to determine the FWHM. b) Fitted FWHM for the PL spectra shown in a). With increasing Ga content (GGI), the FWHM increases. However, the sample with GGI = 0.03 (gray data point), shows a significantly increased FWHM for its Ga content. This is the n-type sample with A > 1. c) Raman spectra of the epitaxial CIGS layers. The n-type absorber with A > 1 exhibits a strongly increased contribution at approximately 152 - 153 $cm^{-1}$ (determined by fitting with a double Lorentzian), which can be ascribed to the ordered defect compound like $CuIn_3Se_5$ [35]. The strong peak a little bit below 180 $cm^{-1}$ corresponds to the $A_1$ mode of CIGS and the small shifts are mainly induced by the different GGI ratios.

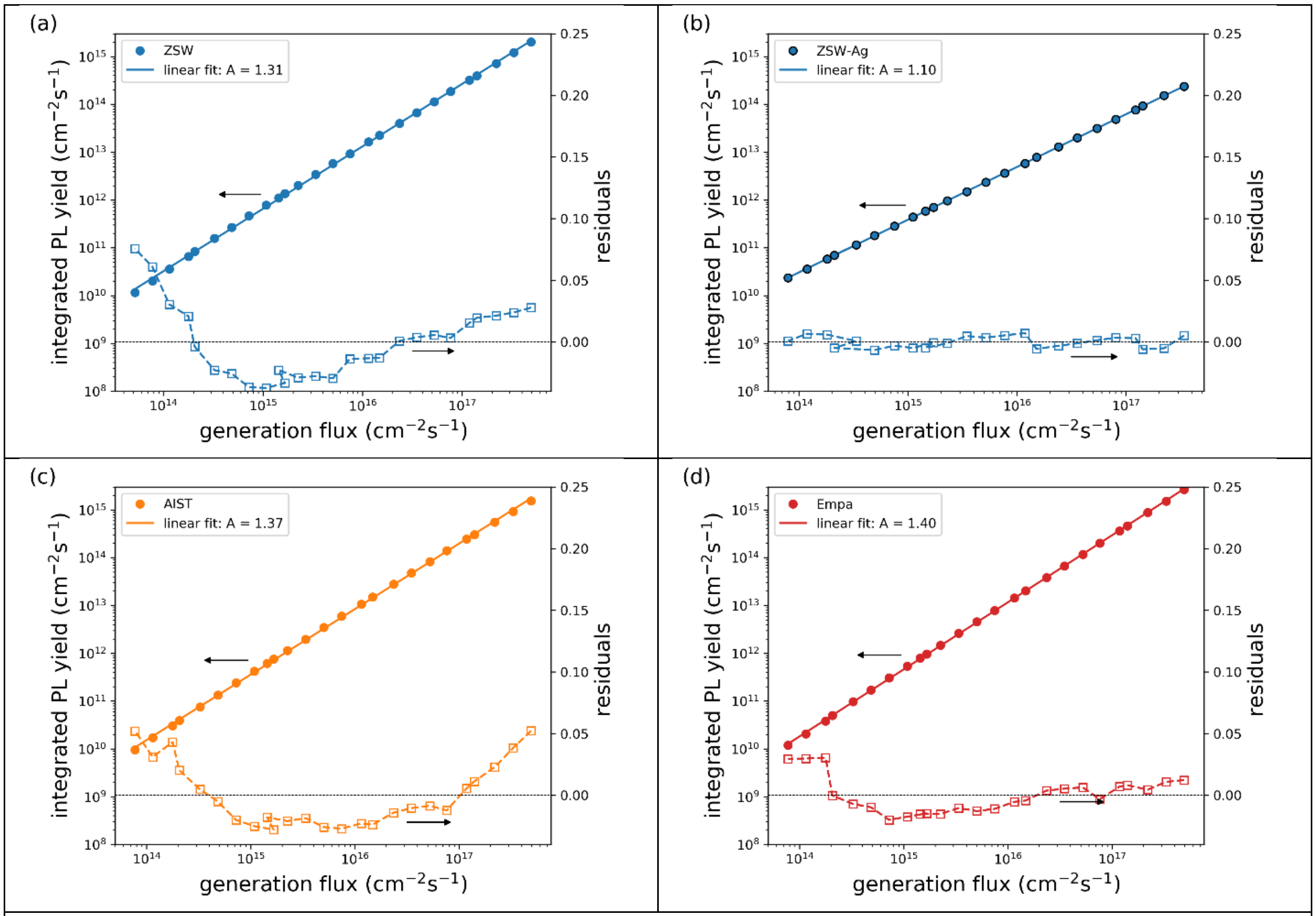

Supplementary Figure 7 – Experimentally measured PL yield (left ordinate) as a function of the generation flux for the four investigated samples shown in

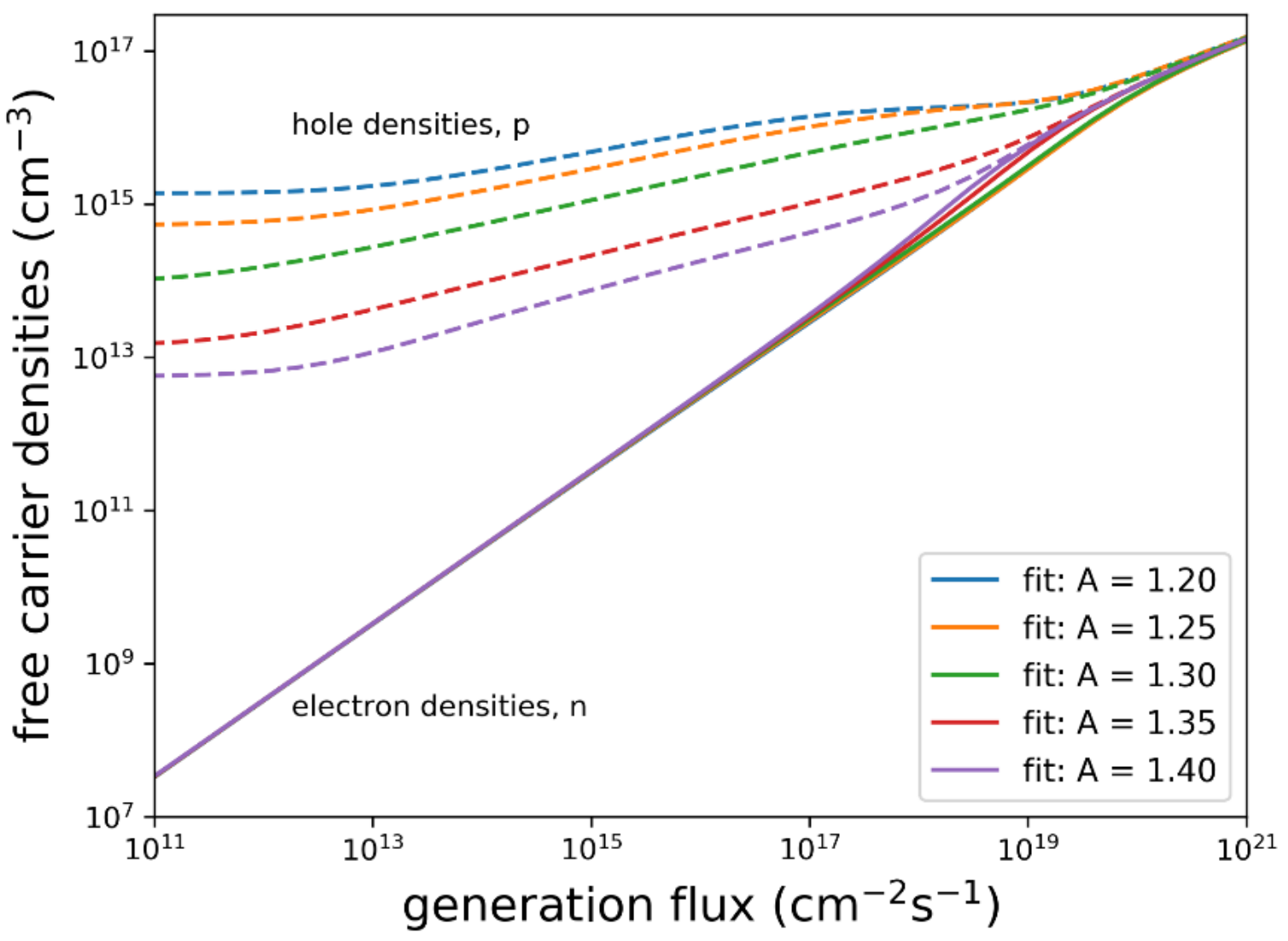

Supplementary Figure 8 – Free electron (solid lines) and hole (dashed lines) densities with respect to the generation flux for the fitted diode factors shown in Figure 1a. The fitting range is between $10^{13}$ and $10^{17}$ cm$^{-2}$s$^{-1}$. Clearly, within in fitting range and even slight above, the semiconductor is in low-injection conditions, i.e. $n \ll p$.

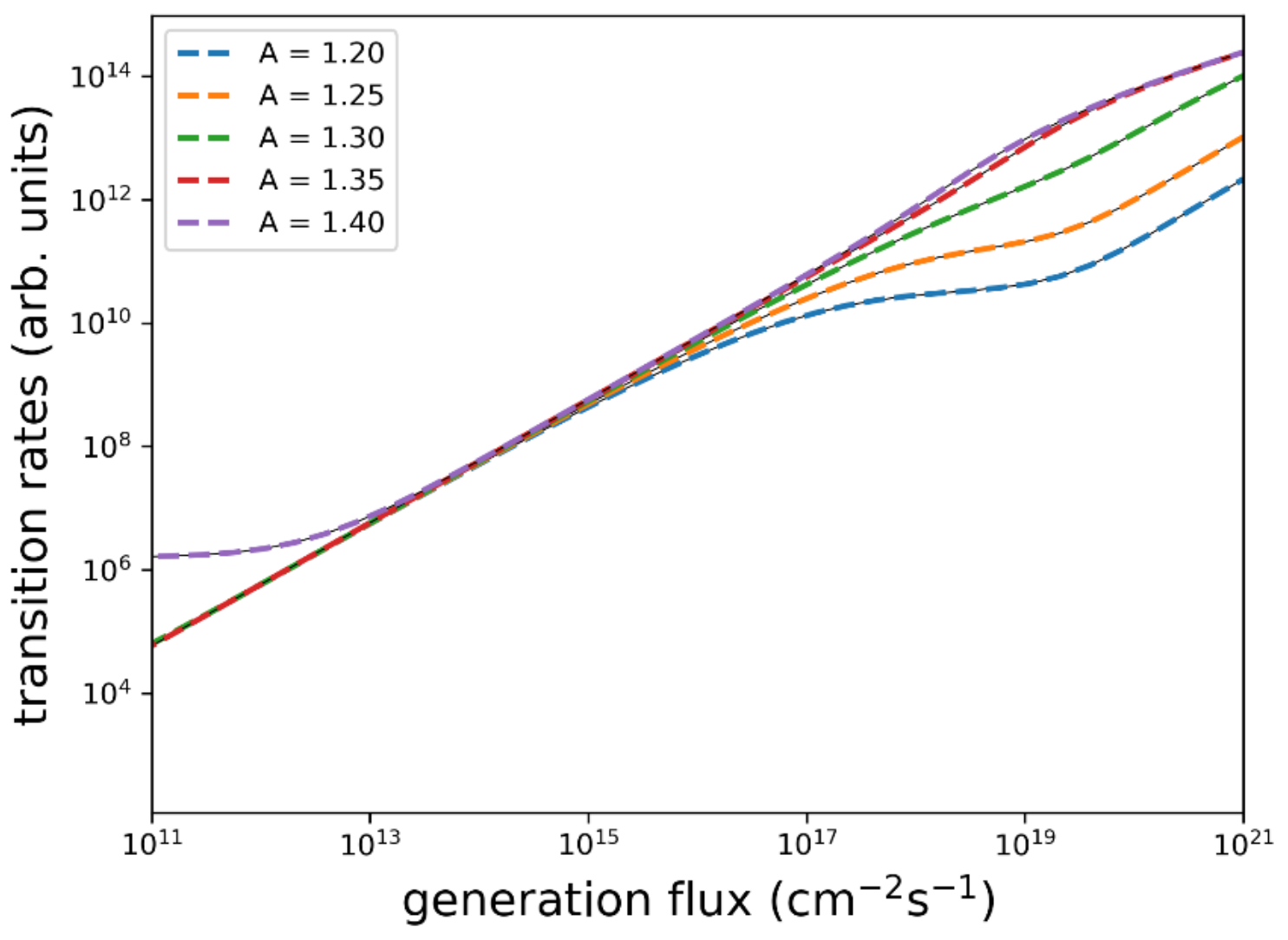

Supplementary Figure 9 – Rates for the acceptor-to-donor (dashed colored lines) and donor-to-acceptor (black solid lines) transitions for the simulations presented in Figure 1 of the main text (see ref. [13] for details). Under steady state conditions, these rates should be equal.

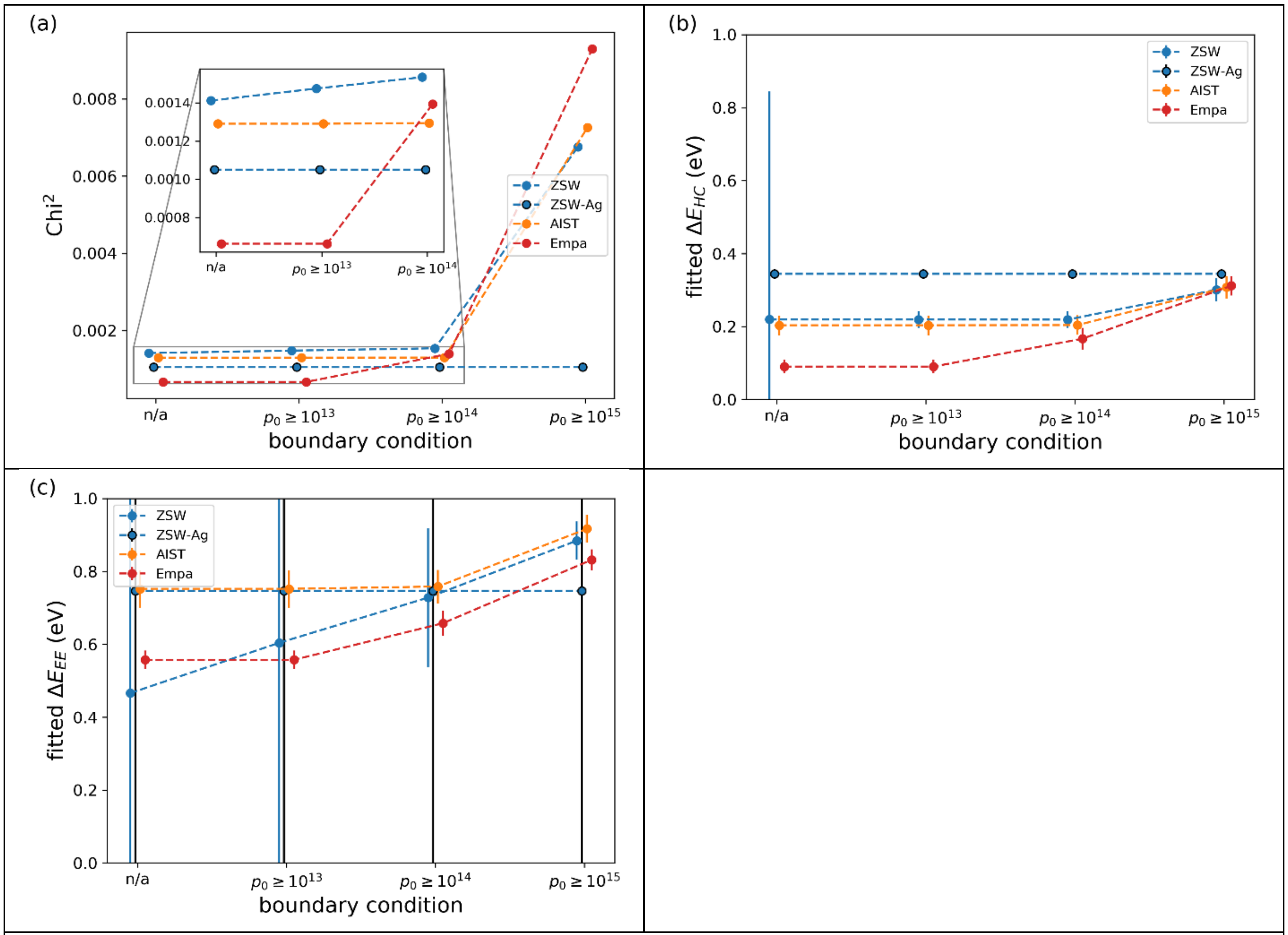

Supplementary Figure 10 – a) Quality of the fit for the four polycrystalline CIGS samples (see Figure 3 and Table 3) as a function of the boundary conditions. The three samples containing Ga throughout the whole absorber film and exhibit emission peaks around 1.10 – 1.15 eV (ZSW, ZSW-Ag, AIST) demonstrate no significant deterioration of the fit quality for the boundary conditions up to $p_{0,bound} = 10^{14} cm^{-3}$. The Empa sample, where Ga is only located towards the back contact, shows an increased $X^2$ of a factor of 2 already for $p_{0,bound} = 10^{14} cm^{-3}$. For $p_{0,bound} = 10^{15} cm^{-3}$, the fit quality deteriorates for all samples with a 'high' diode factor between 1.3 and 1.4. The fit for ZSW-Ag is not impeded by this condition as the best fit already results in a free hole density of $p_0 \approx 6 \times 10^{15} cm^{-3}$. b) and c) the fitted values for the activation energy of the hole capture process $\Delta E_{HC}$ and the electron emission process $\Delta E_{EE}$. Note that the activation energy for the electron capture process is fixed to $\Delta E_{EC} = 0.35\ eV$. Error bars are calculated from the covariance matrix of the fit result. The error bar for $\Delta E_{EE}$ is generally rather large, which indicates that the diode factor is not very sensitive to this parameters. This is a consequence of the rather large values for this energy barrier, which results in a small rate constant $\tau_{EE}^{-1}$ (see Supplementary Figure 3) and thus in a small contribution to the transition from the metastable acceptor to donor state.

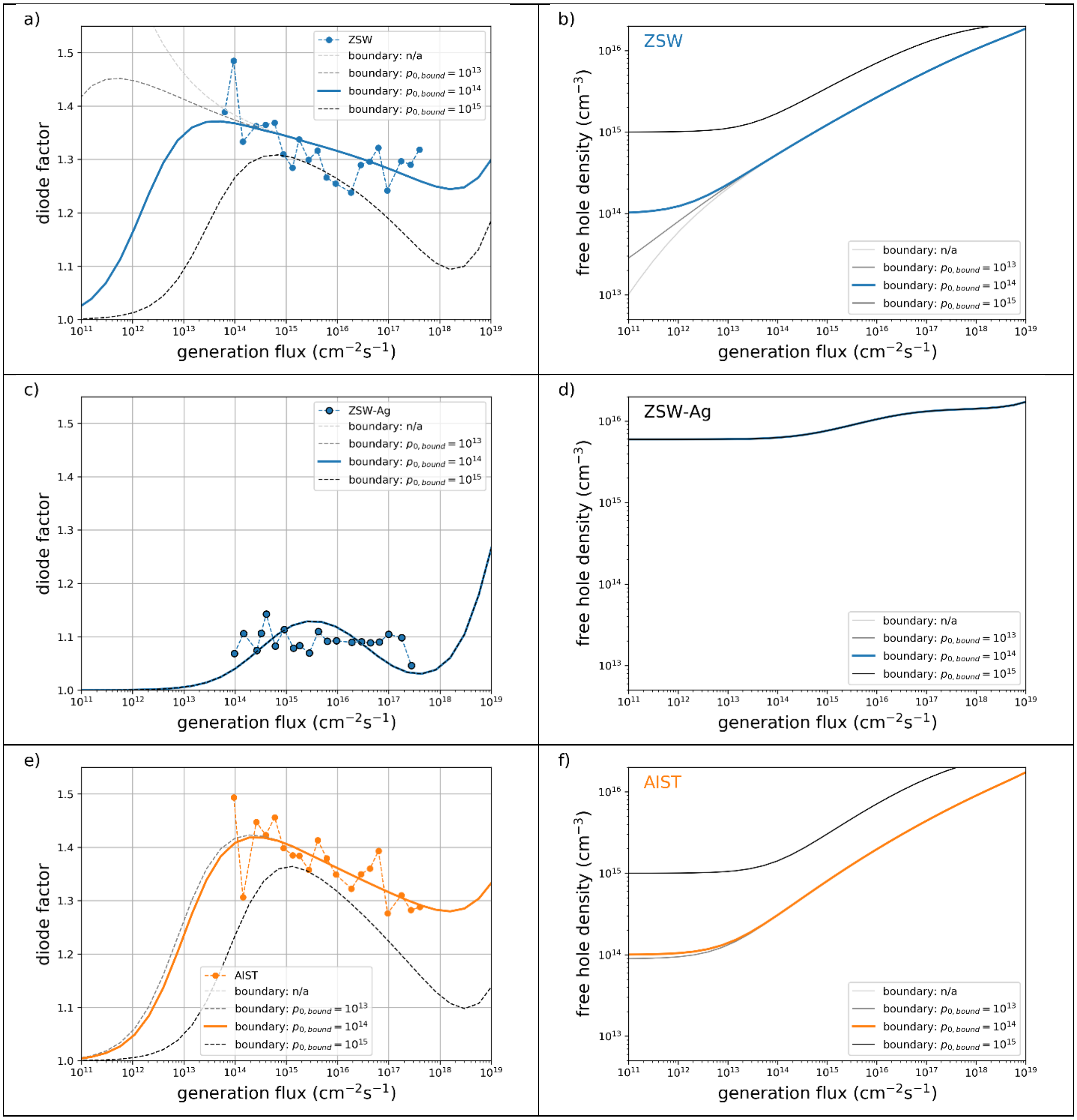

a)
ZSW
boundary: n/a
boundary: $p_{0,bound} = 10^{13}$
boundary: $p_{0,bound} = 10^{14}$
boundary: $p_{0,bound} = 10^{15}$
diode factor
generation flux ($cm^{-2}s^{-1}$)
b)
ZSW
free hole density ($cm^{-3}$)
boundary: n/a
boundary: $p_{0,bound} = 10^{13}$
boundary: $p_{0,bound} = 10^{14}$
boundary: $p_{0,bound} = 10^{15}$
generation flux ($cm^{-2}s^{-1}$)
c)
ZSW-Ag
boundary: n/a
boundary: $p_{0,bound} = 10^{13}$
boundary: $p_{0,bound} = 10^{14}$
boundary: $p_{0,bound} = 10^{15}$
diode factor
generation flux ($cm^{-2}s^{-1}$)
d)
ZSW-Ag
free hole density ($cm^{-3}$)
boundary: n/a
boundary: $p_{0,bound} = 10^{13}$
boundary: $p_{0,bound} = 10^{14}$
boundary: $p_{0,bound} = 10^{15}$
generation flux ($cm^{-2}s^{-1}$)
e)
AIST
boundary: n/a
boundary: $p_{0,bound} = 10^{13}$
boundary: $p_{0,bound} = 10^{14}$
boundary: $p_{0,bound} = 10^{15}$
diode factor
generation flux ($cm^{-2}s^{-1}$)
f)
AIST
free hole density ($cm^{-3}$)
boundary: n/a
boundary: $p_{0,bound} = 10^{13}$
boundary: $p_{0,bound} = 10^{14}$
boundary: $p_{0,bound} = 10^{15}$
generation flux ($cm^{-2}s^{-1}$)

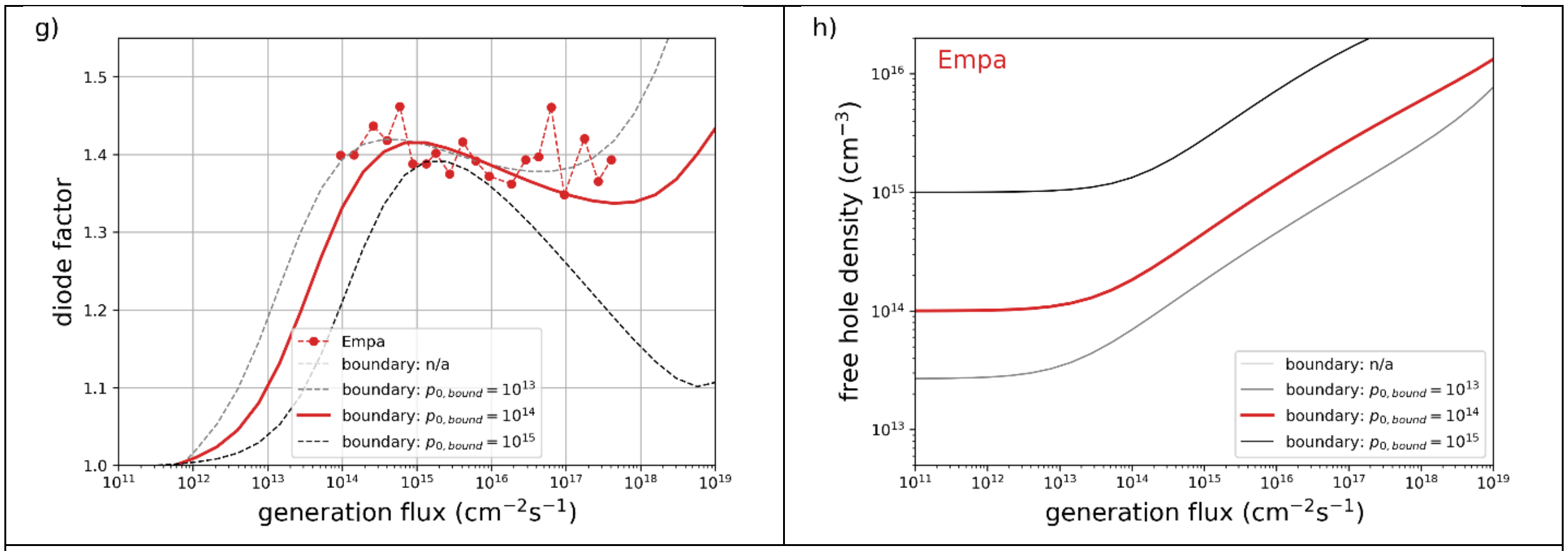

Supplementary Figure 11 – Generation flux dependence of the diode factors measured by PL with fits using the model of metastable defects (a, c, e g) for the four polycrystalline CIGS absorbers (see Figure 3 and Table 3). Different boundary conditions are chosen to ensure a minimum free hole density in the dark (i.e. lowest generation fluxes). The respective generation flux dependence of the free hole density is shown in b, d, f, g. The most stringent boundary of $p_{0,bound} = 10^{15} cm^{-3}$ clearly deteriorates the quality of the fit (see also Supplementary Figure 10 for the $\mathrm{X}^2$ values). Nevertheless, the model is still able to describe the different diode factors. It is noted that only a very simply model is used, i.e. a homogeneous absorber without any bandgap gradient, a homogeneously distributed single SRH midgap recombination center, without surface recombination and a single, homogeneously distributed metastable defect.

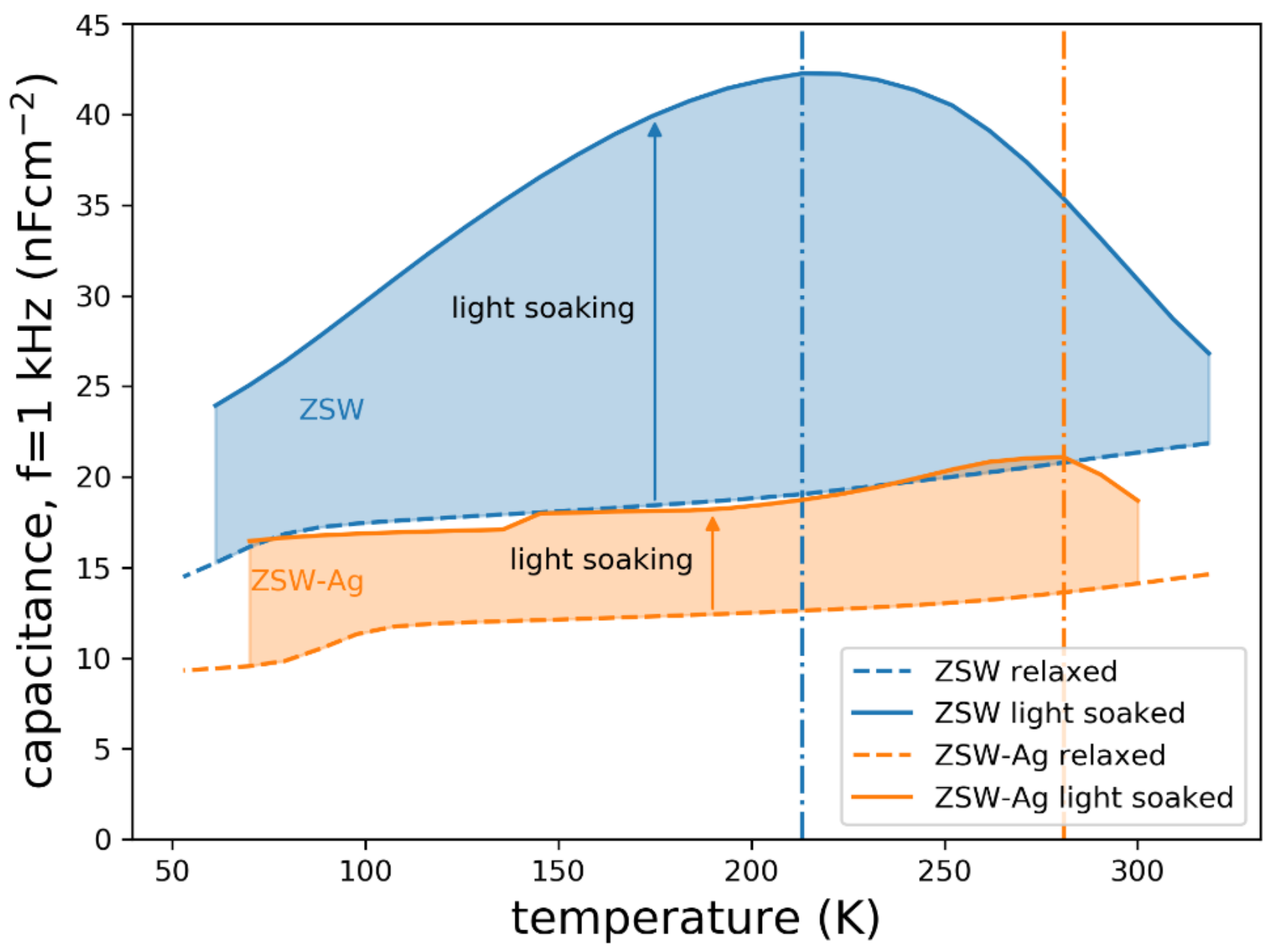

Supplementary Figure 12 – Temperature dependence of the capacitance at an ac-frequency of 1 kHz measured from low to high temperatures. In the relaxed state (dashed lines), the capacitance is continuously increasing with increasing temperature. In the light soaked state, an increased capacitance is observed due to an increased doping density. The capacitance reaches a maximum value at a certain temperature (vertical dash dotted lines). For higher temperatures, the capacitance decreased in the light soaked state, which is attributed to the relaxation of the metastable acceptors into the donor state. For the sample including Ag, this transition is at significantly higher temperatures, which agrees with the higher activation energy $\Delta E_{HC}$ of the hole capture process as determined from the fits of the diode factor (Figure 3 and Table 2).

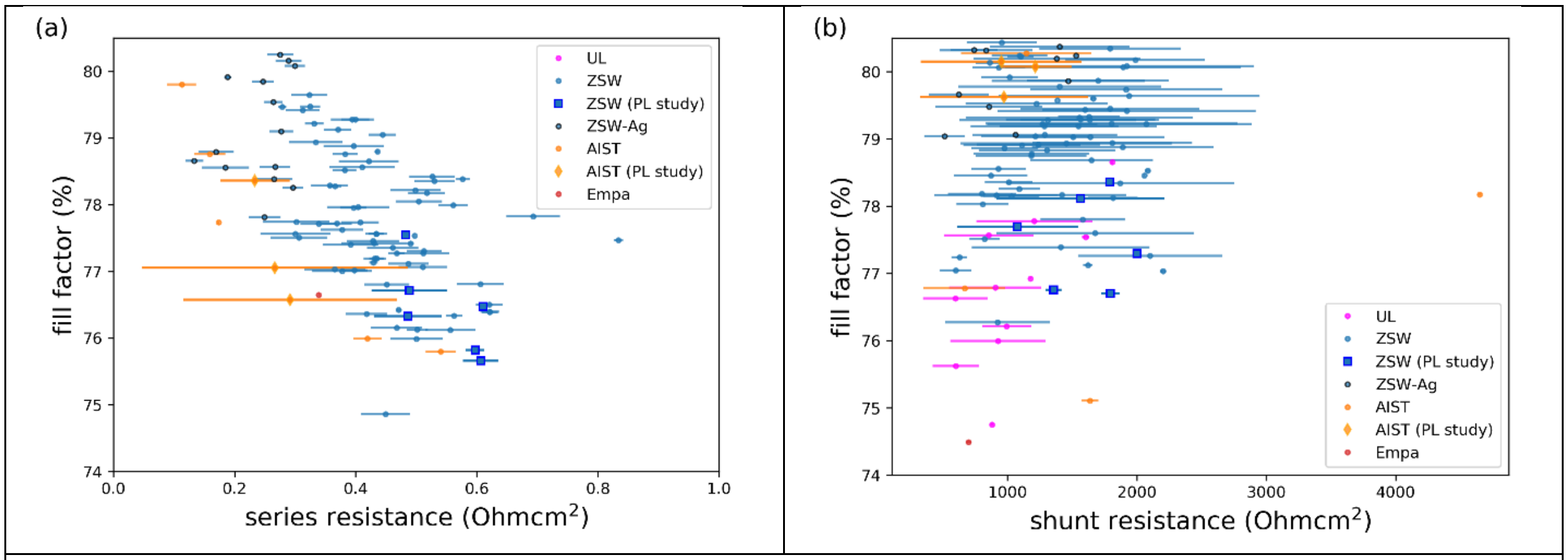

Supplementary Figure 13 – dependence of the fill factor with respect to the series (a) and shunt (b) resistance, extracted with the 1-diode model from illuminated J-V curves.